\documentclass[submission, Phys]{SciPost}
\hypersetup{
    colorlinks,
    linkcolor={red!50!black},
    citecolor={blue!50!black},
    urlcolor={blue!80!black}
}

\usepackage[bitstream-charter]{mathdesign}
\DeclareSymbolFont{usualmathcal}{OMS}{cmsy}{m}{n}
\DeclareSymbolFontAlphabet{\mathcal}{usualmathcal}

\usepackage{graphicx}
\usepackage{hyperref}
\usepackage{xcolor,colordvi}
\usepackage{float}
\usepackage[utf8]{inputenc}
\usepackage{ulem}

\newcommand{\es}[1]{\begin{split}#1\end{split}}
\newcommand{\beq}{\begin{equation}}
\newcommand{\eeq}{\end{equation}}

\newcommand{\aver}[1]{\langle {#1} \rangle}
\newcommand{\lp}{\left(}
\newcommand{\rp}{\right)}
\newcommand{\lsq}{\left[}
\newcommand{\rsq}{\right]}
\newcommand{\lbr}{\left\lbrace}
\newcommand{\rbr}{\right\rbrace}
\newcommand{\da}{\dagger}
\newcommand{\bma}{\begin{pmatrix}}
\newcommand{\ema}{\end{pmatrix}}

\newcommand{\bra}[1]{\langle #1 |}
\newcommand{\ket}[1]{| #1 \rangle}

\newcommand{\rw}{\rightarrow}
\newcommand{\oh}{\frac{1}{2}}

\newcommand{\w}{\omega}
\newcommand{\pt}{\partial _t}

\newcommand{\tr}{\text{tr}}

\newcommand{\abs}[1]{ \left\lvert #1	\right\rvert}

\newcommand{\id}{\mathbb{1}}

\newcommand{\hc}{\text{H.c}}

\newcommand{\mcV}{\hat{\mathcal{V}}}
\newcommand{\mcH}{\hat{\mathcal{H}}}

\begin{document}

% TODO: write your article's title here.
% The article title is centered, Large boldface, and should fit in two lines
\begin{center}{\Large \textbf{
Self-consistent dynamical maps for open quantum systems\\
}}\end{center}

% TODO: write the author list here. Use first name (+ other initials) + surname format.
% Separate subsequent authors by a comma, omit comma and use "and" for the last author.
% Mark the corresponding author with a superscript star.
\begin{center}
Orazio Scarlatella\textsuperscript{1,}\textsuperscript{2},\textsuperscript{3$\star$},
Marco Schirò\textsuperscript{1} 
\end{center}

% TODO: write all affiliations here.
% Format: institute, city, country
\begin{center}
{\bf 1} JEIP, UAR 3573 CNRS, Coll\`{e}ge de France, PSL Research University, 11 Place Marcelin Berthelot, 75321 Paris Cedex 05, France
\\
{\bf 2} Pritzker School of Molecular Engineering, University of Chicago,
5640 South Ellis Avenue, Chicago, Illinois 60637, U.S.A.
\\
{\bf 3} T.C.M. Group, Cavendish Laboratory, J.J. Thomson Avenue, Cambridge CB3 0HE, UK
\\
% TODO: provide email address of corresponding author
${}^\star$ {\small \sf os444@cam.ac.uk}
\end{center}

\begin{center}
\today
\end{center}

% For convenience during refereeing (optional),
% you can turn on line numbers by uncommenting the next line:
%\linenumbers
% You should run LaTeX twice in order for the line numbers to appear.

\section*{Abstract}
{\bf
% TODO: write your abstract here.
In several cases, open quantum systems can be successfully described using master equations relying on Born-Markov approximations, but going beyond these approaches has become often necessary.
In this work, we introduce the NCA and NCA-Markov dynamical maps for open quantum systems, which go
beyond these master equations replacing the Born approximation with a self-consistent approximation, known as non-crossing approximation (NCA). 
These maps are formally similar to master equations, but allow to capture non-perturbative effects of the environment at a moderate extra numerical cost. 
%We also discuss how the NCA maps can be benchmarked by evaluating the leading-order corrections.
To demonstrate their capabilities, we apply them to the spin-boson model at zero temperature for both a Ohmic and a sub-Ohmic environment, showing that they can both qualitatively capture its strong-coupling behaviour, and be quantitatively correct beyond standard master equations.
}

%% TODO: include a table of contents (optional)
%% Guideline: if your paper is longer that 6 pages, include a TOC
%% To remove the TOC, simply cut the following block
%\vspace{10pt}
%\noindent\rule{\textwidth}{1pt}
%\tableofcontents\thispagestyle{fancy}
%\noindent\rule{\textwidth}{1pt}
%\vspace{10pt}

\section{Introduction}
The theory of open quantum systems, born to describe nuclear magnetic resonance (NMR) \cite{wangsnessBloch1953,redfieldRedfield1957,kuboKubo1969} and lasers \cite{scullyLamb1967,mollowMiller1969}, is now of fundamental importance for the development of quantum devices~\cite{mccauleyJacobs2020,leghtasDevoret2015,giovannettiMaccone2011,vinjanampathyAnders2016a,
komarLukin2014,gehringvanderZant2019}, to describe chemical reactions \cite{hanggiBorkovec1990,wolynesWolynes1981,vothMiller1989,ianconescuPollak2019}, understand biological complexes \cite{engelFleming2007,panitchayangkoonEngel2010,colliniScholes2010,blankenshipSayre2011,
lambertNori2013,caoZigmantas2020} and to explore novel non-equilibrium quantum states~\cite{kapitSimon2014a,maSchuster2019,puertasmartinezRoch2019a,carusottoSimon2020a}.

For typical quantum optical systems, where the coupling with the environment is weak and the environment is unstructured, one can rely on Born-Markov master equations~\cite{scullyZubairy1997,redfieldRedfield1957,daviesDavies1974,lindbladLindblad1976a,dumckeSpohn1979,nathanRudner2020,mozgunovLidar2020,mccauleyJacobs2020a,
davidovicDavidovic2020,trushechkinTrushechkin2021,becker2021lindbladian} and on equivalent stochastic approaches \cite{dalibardMolmer1992,dumRitsch1992,
gardinerZoller2004}. 
%In many relevant cases for quantum optics \cite{scullyZubairy1997}, quantum information \cite{nielsenChuang2010} and ultracold atoms \cite{syassenDurr2008a} one can rely on weak-coupling Markovian master equations~\cite{redfieldRedfield1957,daviesDavies1974,lindbladLindblad1976a,dumckeSpohn1979,nathanRudner2020,mozgunovLidar2020,mccauleyJacobs2020a,
%davidovicDavidovic2020,trushechkinTrushechkin2021,becker2021lindbladian} and on equivalent stochastic approaches \cite{dalibardMolmer1992,dumRitsch1992,
%gardinerZoller2004}.
Nevertheless, nowadays going beyond these approaches is necessary for a growing number of cases, including electronic transport problems~\cite{espositoGalperin2009,espositoGalperin2010,
jinYan2014,sowaGauger2018,sowaGauger2020},
%quantum optical systems \cite{hoeppeBusch2012,gonzalez-tudelaCirac2017, gonzalez-tudelaCirac2017a, gonzalez-tudelaCirac2018}, 
optomechanical resonators~\cite{groblacherEisert2015}, quantum dots~\cite{madsenLodahl2011,miPetta2017}, superconducting circuits~\cite{whiteModi2020,papicdeVega2023,paladinoAltshuler2014,burnettBylander2019,rowerOliver2023} and
quantum simulation platforms~\cite{hansonAwschalom2008, lebratEsslinger2018, liuPiilo2011, maierRoos2019, recatiZoller2005, puertasmartinezRoch2019} 
%and in quantum information \cite{mccauleyJacobs2020},
%
%including relevant for condensed matter physics \cite{groblacherEisert2015}, quantum information \cite{mccauleyJacobs2020}, quantum optics \cite{hoeppeBusch2012,gonzalez-tudelaCirac2017, gonzalez-tudelaCirac2017a, gonzalez-tudelaCirac2018} and quantum simulation~ \cite{haikkaManiscalco2011, hansonAwschalom2008, lebratEsslinger2018, liuPiilo2011, maierRoos2019, puertasmartinezRoch2019, recatiZoller2005,magazzuGrifoni2018},
where 
% the system-bath coupling is strong and therefore 
non-Markovian and non-perturbative phenomena are expected~\cite{devega2017dynamics,liPiilo2020,
maierRoos2019,chinPlenio2012,
alickiHorodecki2002,dongSun2018}.

%for stronger system-bath couplings, as it is the case for several systems which are currently investigated in condensed matter physics \cite{groblacherEisert2015}, quantum information \cite{mccauleyJacobs2020}, quantum optics \cite{hoeppeBusch2012,gonzalez-tudelaCirac2017, gonzalez-tudelaCirac2017a, gonzalez-tudelaCirac2018} and quantum simulation~ \cite{haikkaManiscalco2011, hansonAwschalom2008, lebratEsslinger2018, liuPiilo2011, maierRoos2019, puertasmartinezRoch2019, recatiZoller2005,magazzuGrifoni2018}, where non-Markovian dynamics~\cite{devega2017dynamics,flanniganDaley2021,maierRoos2019, magazzuGrifoni2018,chinPlenio2012,
%alickiHorodecki2002,dongSun2018,liPiilo2020} and other strong-coupling phenomena are expected.

Theoretical approaches beyond Born-Markov master equations have recently been developed, including phenomenological master equations~\cite{shabaniLidar2005,maniscalcoPetruccione2006,
chruscinskiKossakowski2010a,espositoGalperin2009,
espositoGalperin2010,jinYan2014}, but also microscopic approaches such as
diagrammatic techniques \cite{schoellerSchon1994,schoellerSchoeller2009,karlewskiMarthaler2014,mullerStace2017,lindnerSchoeller2018,
kleinherbersSchutzhold2020,mulhbacher2008realtime} including diagrammatic Monte-Carlo \cite{schiro2009realtime,werner2009diagrammatic,cohen2014green,
cohen2015taming,chenReichman2017b,
chenReichman2017g}, hierarchies of exact equations of motion \cite{jinYan2008,hartleMillis2013,tanimuraTanimura2020,diosiStrunz1997,
diosiStrunz1998,gaspardNagaoka1999,jingYu2012,suessStrunz2014} and matrix product states approaches \cite{thoennissAbanin2022,strathearnLovett2018}.
The latter microscopic techniques can be very powerful, but they are often computationally demanding and formally more involved than Born-Markov master equations, limiting their applicability in many cases.

On the other hand, a powerful technique of many-body theory is performing partial summations of perturbation series, such as in several textbook applications \cite{altland_simons_2010,bruusFlensberg2004}. 
This allows to obtain analytical equations that, despite being simple, can capture complex non-perturbative phenomena.
% \new{NEED RM}On the other hand, in a different context such as in textbook applications of many-body theory\cite{bruusFlensberg2004}, often very simple non-perturbative approaches have been devised in the form of self-consistent approximations corresponding to partial resummations of perturbation theories, that can capture strong-coupling phenomena but still rely on simple equations. 
An example are the non-crossing approximations (NCA), that have been used for example for disordered systems~\cite{altland_simons_2010}, for quantum impurity models, both in and out of equilibrium~\cite{bickersBickers1987b,muller-hartmannMuller-Hartmann1984,
nordlander1999how,ecksteinWerner2010a,
chen2016anderson}, and for quantum transport problems
\cite{meir1993low,pruschkeJarrell1993,
erpenbeck2021revealing,hartle2013decoherence}.
In \cite{schiroScarlatella2019} a NCA has been formulated 
to conveniently capture the dynamics of quantum systems coupled simultaneously to a Markovian and to a non-Markovian environment. 

%In \cite{schiroScarlatella2019}, a NCA has been developed to compute the reduced dynamics of quantum systems coupled simultaneously to a Markovian and to a non-Markovian environment. 
%, as it naturally arises in the context DMFT problems \cite{scarlatellaSchiro2021b}.

In this work, we show how for generic open quantum systems the non-crossing approximation of \cite{schiroScarlatella2019} can be used to go beyond the Born approximation, underlying standard master equations such as the Redfield \cite{redfieldRedfield1957} or Lindblad-Davies \cite{daviesDavies1974,lindbladLindblad1976a}, and that it can also be combined with a usual Markovian approximation. %, resulting in a time-local equation. 
%\new{In this work, we show how for generic open quantum systems one can go beyond standard master equations by replacing the Born approximation with the NCA introduced in \cite{schiroScarlatella2019}. }
%\new{In this work, we show how for generic open quantum systems this approximation leads to a simple generalization of standard master equations, where the Born approximation is replaced by the NCA. }
%\new{In this work, we consider instead a conventional open quantum system setup and show how the same NCA corresponds to replacing the Born approximation \cite{breuerPetruccione2007}, underlying standard weak coupling master equations, with a self-consistent Born approximation.}
%that this NCA can be used to replace the Born approximation \cite{breuerPetruccione2007}, underlying standard weak coupling master equations. %
The result is an equation for the dynamical map propagating the system density matrix, rather than for the density matrix itself, that we call the NCA or NCA-Markov map, which is formally very similar and reduces to the usual Born and Born-Markov master equations at sufficiently weak coupling, but has a \textit{self-consistent} dissipator allowing to capture non-perturbative effects at a moderate extra numerical cost. 
Furthermore, we discuss how the NCA maps can be benchmarked evaluating the leading-order self-consistent correction, the one crossing approximation (OCA).
%, and that these approximations can both yield 
%qualitatively correct results at strong system-bath coupling and
%quantitatively correct results in the weak-coupling regime, beyond the capabilities of standard master equations. 

%Furthermore, we discuss how leading order corrections can be evaluated, to asses the validity of the NCA maps predictions.
We then apply the NCA and NCA-Markov maps to the spin-boson model with a zero temperature environment. 
We show that in the case of a Ohmic environment both approaches correctly capture the non-perturbative dynamics of this model at strong system-bath coupling, including its crossover from coherent to incoherent dynamics and its quantum phase transition to a localized phase, which are completely missed by the Born and Born-Markov master equations. 
Finally, computing the one-crossing correction to NCA we show for both a Ohmic and a sub-Ohmic environment that these approaches are quantitatively accurate in regimes in which the latter master equations display significant deviations.

\section{From master equations to dynamical maps}  
\label{sec:fromBornToNCA}

We consider a generic quantum system coupled to a bath with total Hamiltonian \\ $ H_{\rm tot }= H_S + H_B + H_{SB} $, where $H_S$ and $H_B$ are the system and bath Hamiltonians. 
The main assumptions behind the self-consistent maps are that the bath is described by a set of non-interacting bosonic or fermionic modes and that the system-bath density matrix factorizes at time $t=0$, $\rho_{\rm tot} (0) = \rho (0) \otimes \rho_B(0) $. 
To introduce the method, we restrict here to a single bath of bosonic modes, with annihilation operators $a_{i}$ and Hamiltonian $H_B = \sum_{i} \w_{i} a_{i}^\da a_{i}$, and to a system-bath coupling of the form $H_{SB} = X 
\otimes B$, respectively with system and bath operators $X = X^\da$ and $B = \sum_i \frac{ \lambda_{i}}{2} ( a_{i} + a^\da_{i} ) = B^\da$ and coupling strength $\lambda_{i}$. We also consider an initial bath density matrix $\rho_B(0)$ that is stationary with respect to $H_B$.  We remark though that the maps apply in more general cases, as we discuss in Sec. \ref{sec:beyondNCA}.

Under those assumptions, the reduced dynamics of the system is often conveniently described by master equations of the form \cite{breuerPetruccione2007} 
\begin{align}
\label{eq:BornEq}
\partial_{t} \rho(t) = \hat{\mathcal{H}}_S \rho(t) + \int_{0}^{t} d t_1 \hat{\mathcal{D}}(t-t_1) \rho(t_1) 
\end{align}
where $\hat{\mathcal{D}}$ is the dissipator, describing the influence of the bath on the system. 
Weak-coupling master equations are based on a second-order approximation in the system-bath coupling, the Born approximation \cite{breuerPetruccione2007}, that 
%assuming a factorized system-bath density matrix at time $t=0$, $\rho_{\rm tot} (0) = \rho (0) \otimes \rho_B(0) $, and a bath density matrix $\rho_B(0)$ that is stationary with respect to $H_B$, 
leads to the dissipator 
%leads to the Born master equation 
%\begin{align}
%\label{eq:BornEq}
%\partial_{t} \rho(t) = \hat{\mathcal{H}}_S \rho(t) + \int_{0}^{t} d t_1 \hat{\mathcal{D}}_{\rm Born}(t-t_1) \rho(t_1) 
%\end{align}
%where $\hat{\mathcal{D}}_{\rm Born}(\tau)$ is the Born dissipator 
\beq
\label{eq:bornSe} 
\hat{\mathcal{D}}_{\rm Born}(\tau) = \Gamma(\tau) \lp  e^{\mcH_s \tau } \lsq X \bullet  \rsq X    -   X e^{\mcH_s \tau } X \bullet \rp + \hc 
\eeq
in which the superoperator $\mcH_s \bullet = - i \lsq H_S , \bullet \rsq $ describes the bare system dynamics and where $\Gamma$ is the bath correlation function 
\beq 
\label{eq:bathCorr}
\Gamma (\tau) = \tr \lsq B(\tau) B (0) \rho_B(0) \rsq 
\eeq 
with $B(\tau)$ time-evolved with the bath Hamiltonian $H_B$. We used square brakets in \eqref{eq:bornSe} when necessary to indicate the argument of superoperators. % and we remark that no Markovian approximation has been done at this point. 
We refer to \eqref{eq:BornEq}, \eqref{eq:bornSe} as the Born master equation. 
The dissipator \eqref{eq:bornSe} has the familiar structure of Redfield \cite{redfieldRedfield1957} and Lindblad \cite{lindbladLindblad1976a} master equations, as it is the starting point for deriving them. 
% as the Born master equation defined by \eqref{eq:BornEq} and \eqref{eq:bornSe} is the starting point for deriving them. 
%The Born master equation defined by \eqref{eq:BornEq} and \eqref{eq:bornSe} is the starting point for deriving Redfield \cite{redfieldRedfield1957} and Lindblad-Davies \cite{lindbladLindblad1976a} equations and its dissipator. 

Heuristically, we introduce the NCA self-consistent map in two steps:
%Heuristically, one can promote the Born approximation to a self-consistent Born approximation in two steps. 
first we recognize that the dynamical map $\mcV(t)$, namely the superoperator evolving the system density matrix from the initial state, $\rho(t) = \mcV(t) \rho(0)$, obeys an analogous equation:
\beq
\label{eq:dyson}
\partial_{t} \hat{\mathcal{V}}(t)=\hat{\mathcal{H}}_S \hat{\mathcal{V}}(t)+\int_{0}^{t} d t_{1} \hat{\mathcal{D}}\left(t-t_{1}\right) \hat{\mathcal{V}}\left(t_{1}\right)
\eeq
Note that such a map can always be defined, and \eqref{eq:dyson} only relies on the two main assumptions outlined above. 

Then, we make the Born dissipator self-consistent, by replacing in \eqref{eq:bornSe} the evolution superoperator $e^{\mcH_s \tau}$, describing the dynamics of the system isolated from the environment, with 
the dynamical map $\mcV(\tau)$ itself, taking into account the influence of the environment:
\beq
\label{eq:ncaSe} 
\hat{\mathcal{D}}_{\rm NCA}(\tau) = \Gamma(\tau) \lp  \mcV(\tau) \lsq X \bullet  \rsq X    -   X \mcV(\tau) X \bullet \rp + \hc 
\eeq
where for taking the Hermitian conjugate one can use the property $( \mcV [ \bullet ] )^\dagger  = \mcV[ \bullet^\da ]$. 
%The dynamical map $\mcV$ solving Eq. \eqref{eq:dyson} with dissipator \eqref{eq:ncaSe} is the NCA dynamical map. 
%Turning bare equations into self-consistent ones is a common trick in field theoretical approaches to obtain non-perturbative approximations, that might be able to qualitatively capture the physics beyond the perturbative regime.
Eq. \eqref{eq:dyson} with the dissipator \eqref{eq:ncaSe} has a very similar structure to standard master equations, but is an equation for the dynamical map rather than for the density matrix. We call its solution the \textit{NCA dynamical map}. 
Differently from standard master equations, the dissipator \eqref{eq:ncaSe} is determined \textit{self-consistently} as it depends on the unknown. 
Note also that \textit{non-Markovian effects} are captured both by the time-integral in Eq. \eqref{eq:dyson} and by the self-consistent dissipator, coupling the map to its values at earlier times.
%The NCA dynamical map is the solution of Eq. \eqref{eq:dyson} solved jointly with Eq. \eqref{eq:ncaSe} for the dissipator, that is determined ``self-consistently'' as it depends on the unknown.  

\begin{figure}[h]
\center
\includegraphics[width=0.6\linewidth]{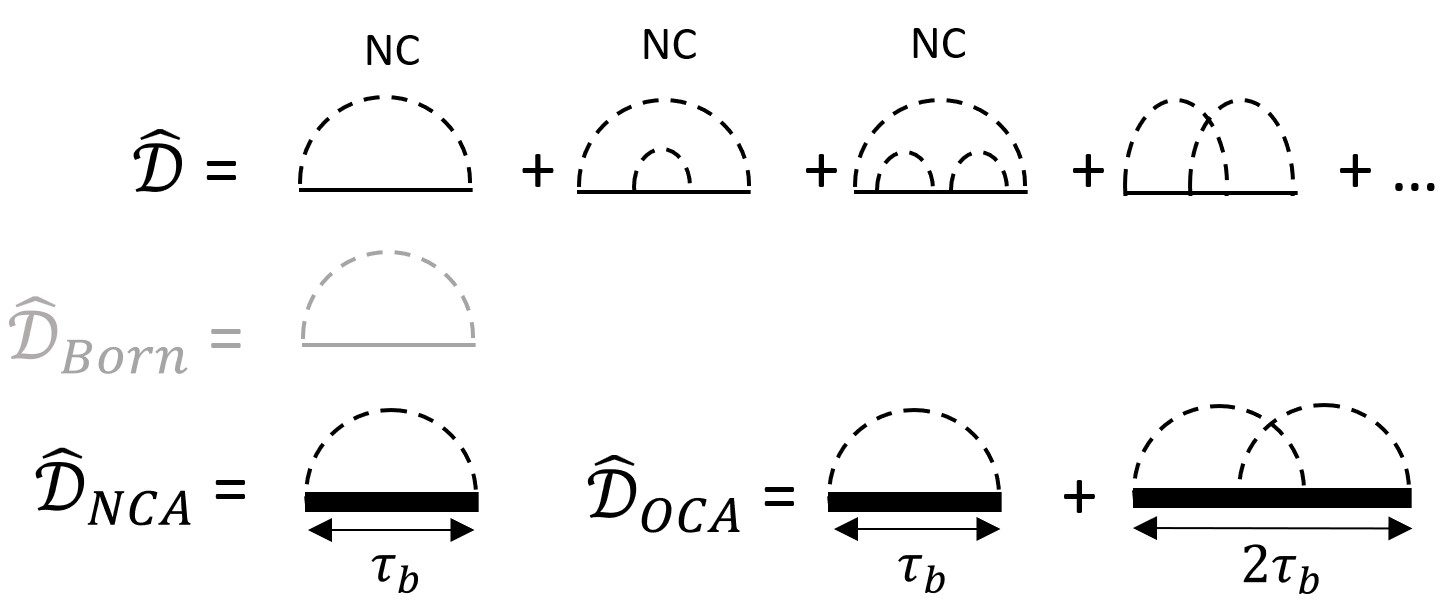}
\caption{ The expression at the top represents the (first terms of the) exact series for the dissipator $\hat{\mathcal{D}}$ which is the self-energy of the Dyson equation \eqref{eq:dyson}: non-bold solid lines represent bare evolution superoperators $e^{\mcH_s (\tau)}$, dashed lines correspond to bath correlation functions \eqref{eq:bathCorr}, with a typical decay time $\tau_b$. %(see \cite{Note1}).  
The middle expression shows that retaining the 1st term of the series corresponds to the Born dissipator \eqref{eq:bornSe}.
At the bottom, the NCA dissipator \eqref{eq:ncaSe} corresponds to the sum of all the diagrams in which the dashed lines do not cross (labelled by ``NC''), which can be expressed in terms of a bold solid line representing the dynamical map $\mcV$. 
Similarly, $\hat{\mathcal{D}}_{\rm OCA}$ is the dissipator including the leading-order self-consistent correction to NCA, namely in the one-crossing approximation (OCA). 
We also show that $\hat{\mathcal{D}}_{\rm OCA}$ decays on a longer time-scale $2\tau_b$ than $\hat{\mathcal{D}}_{\rm NCA}$, decaying in $\tau_b$.
%Finally, we remark that the NC diagrams decay on the timescale $\tau_b$ in which bath correlation functions decay, while the exact $\hat{\mathcal{D}}$ contains also diagrams decaying on longer time-scales. 
}
\label{fig:seDiag}
\end{figure}

Apart from this heuristic derivation, a formal derivation using diagrammatic techniques is given in App. \ref{app:formalDer}, where \eqref{eq:dyson} is recognized to be the Dyson equation for the perturbation theory of the dynamical map $\mcV(t)$ in the system-bath coupling and the dissipator $\hat{\mathcal{D}}$ to be its self-energy (the same is true for Eq. \eqref{eq:BornEq} and $\rho(t)$).
An important consequence is that \eqref{eq:dyson} (as \eqref{eq:BornEq}) is in principle an exact equation for bosonic or fermionic environments, where the exact self-energy $\hat{\mathcal{D}}$ is defined by its perturbation series.
A diagrammatic representation of this series along with the approximations schemes considered in this manuscript is shown in Fig. \ref{fig:seDiag}: a lowest-order truncation of the series corresponds to the Born approximation \eqref{eq:bornSe}, while approximating it with the sum of its infinitely-many ``non-crossing'' diagrams (hence the name NCA) one finds that the sum of the series has a compact expression in terms of $\mcV(t)$, namely Eq. \eqref{eq:ncaSe}.

The NCA map is non-perturbative, as its dissipator \eqref{eq:ncaSe} includes contributions to all orders in the bath correlation function $\Gamma$ (or in the system-bath coupling $H_{SB}$), in contrast with the Born dissipator \eqref{eq:bornSe} which is first order in $\Gamma$ (2nd order in $H_{SB}$). 
%To show how the NCA map is non-perturbative, we remark that in contrast with the Born dissipator \eqref{eq:bornSe} that is first order in the bath correlation function $\Gamma$ (or equivalently second order in the system-bath coupling $H_{SB}$), the NCA dissipator \eqref{eq:ncaSe} has contributions to all orders in $\Gamma$.
To see this explicitly, one can obtain ${\mcV}$ by integrating Eq. \eqref{eq:dyson} with dissipator \eqref{eq:ncaSe}, and plug it back in Eq. \eqref{eq:ncaSe} to get an expression for the dissipator which explicitely depends on $\Gamma$, $\Gamma^2$ and ${\mcV}$: substituting ${\mcV}$ recursively, contributions up to all powers in $\Gamma$ are generated. 
This allows the NCA map to capture non-perturbative effects with respect to standard master equations based on the Born approximation, like Redfield \cite{redfieldRedfield1957} and Lindblad-Davies equations \cite{daviesDavies1974,lindbladLindblad1976a}.
We also remark that, when the coupling is sufficiently weak the NCA maps become equivalent to standard weak-coupling master equations, as to lowest order in the system-bath coupling 
the NCA dissipator \eqref{eq:ncaSe} reduces to the Born one \eqref{eq:bornSe}: ${\mcV}(\tau)$ in \eqref{eq:ncaSe} reduces to $e^{\mcH_s \tau} $.

A discussion of the regimes of validity of the NCA maps, of how to benchmark their predictions and of their applicability is 
% and on how to benchmark their predicitions introducing the leading-order self-consistent corrections is 
reported in Section \ref{sec:beyondNCA}. 
In the following Section we show how the usual Markovian approximation done for master equations can be combined with the NCA, how a similar steady-state equation holds and also a quantum regression theorem for computing correlation functions.  
% while other points in common with those equations are the steady-state equation reported in \ref{sec:steadyState} and a quantum regression theorem for computing correlation functions in \ref{app:corrFunc}. 

%\new{Finally, we note that earlier attempts towards a self-consistent Born approximation for master equations exist in the literature \cite{espositoGalperin2009,espositoGalperin2010,
%jinYan2014}, but differently from the present work,
%those do not achieve a fully self-consistent equation in which the same dynamical map evolving the system density matrix appears in the dissipator, but they rather propose a partially self-consistent approach in which a separate equation is used to obtain the dissipator, which therefore does not retain contributions to all-orders in the system-bath coupling and for which our claims in Sec. \ref{sec:beyondNCA} regarding the validity and generalizations of our approach are invalidated. }

%Apart from the heuristic derivation presented here, a formal derivation of the NCA equations using diagrammatic techniques is given in App. \ref{app:formalDer}, where equation \eqref{eq:dyson} is recognized to be the Dyson equation for the perturbation series of the dynamical map $\mcV$ in the system-bath coupling and the dissipator $\hat{\mathcal{D}}_{NCA}$ \eqref{eq:ncaSe} plays the role of its self-energy in the NCA. 
%The formal derivation is particularly useful to have a picture of the physical processes included in this approximation and to generalize it to higher-order self-consistent schemes useful for benchmarking.

\subsection{Markovian approximation, steady-state and quantum regression theorem} 

When the bath-induced dynamics on the system is slower than the typical decay time of the dissipator, one can do a partial \textit{Markovian approximation}, replacing $\mcV(t_1)$ in \eqref{eq:dyson} with $e^{\mcH_s (t_1-t)} \mcV(t) $, yielding: 
\beq
\label{eq:markDys}
\partial_{t} \hat{\mathcal{V}}(t)= \lsq \hat{\mathcal{H}}_S +\int_{0}^{t} d \tau \hat{\mathcal{D}}\left(\tau \right)  e^{ - \mcH_s (\tau)}  \rsq \hat{\mathcal{V}}\left(t\right)
\eeq
We call the solution of \eqref{eq:markDys} and \eqref{eq:ncaSe} the NCA-Markov dynamical map. 
The same approximation is routinely done at the level of master equations \eqref{eq:BornEq}, where $\rho(t_1)$ is replaced with $e^{\mcH_s (t_1-t)} \rho(t)$, leading to what we will refer to as the Born-Markov master equation, that we will use for comparison.  
The NCA-Markov has a computational advantage over the NCA, as the integral in \eqref{eq:markDys} can be updated at each time-step, and does not need to be recomputed entirely as in \eqref{eq:dyson}, leading to a reduced numerical cost as we will discuss in Sec. \ref{sec:numericalCost}.
We remark that this approximation does not lead to a fully Markovian equation. This is already true for the Born-Markov equation we will compare with, where the finite upper time-integration limit, analogously as in  \eqref{eq:markDys}, introduces some non-Markovian effects and results in a more accurate approach \cite{hartmannStrunz2020} (see also Appendix \ref{app:weakCoupME}). The NCA-Markov map captures even further non-Markovian effects, as it still depends explicitly on its previous times through the self-consistent dissipator \eqref{eq:ncaSe}. 
We further discuss the validity of the NCA-Markov approximation in App. \ref{app:validityNCAMarkov}.

\label{sec:steadyState}

For master equations, the \textit{steady-state} can be computed by the the generator of their infinitesimal time-evolution, the Liouvillian, by an algebraic equation. An analogus expression exists here. 
Assuming that the system forgets its initial conditions at a sufficiently long time and reaches a time-independent steady-state, defined as $ \rho_s = \lim_{t\rw \infty} \mcV(t,0) \rho(0) $, this state also obeys the equation \cite{schiroScarlatella2019,scarlatellaSchiro2021b}:
\beq
\label{eq:steadyState}
\left(\hat{\mathcal{H}}_S+\int_{0}^{\infty} d \tau \hat{\mathcal{D}}(\tau)\right) \rho_{s}=0
\eeq
We remark that at weak system-bath coupling $\hat{\mathcal{D}}(\tau)$ is expected to decay on a much shorter timescale than that for the system to reach the steady-state. Eq. \eqref{eq:steadyState} then allows to extract $\rho_s$ from the short-time dynamics of the system, akin to the condition for Markovian master equations. 

% akin to the condition for finding the steady-states of Markovian master equations in terms of the generator of their infinitesimal time-evolution, the Liouvillian.

% \section{Steady-state correlation functions}
\label{app:corrFunc}

\textit{The quantum regression theorem} for Markovian master equations allows to compute (multi-time) correlation functions from the same generator of the dynamics which evolves the density matrix \cite{breuerPetruccione2007}. 
This generalizes to NCA, for which it can be proven without making a Markovian assumption \cite{scarlatellaSchiro2021b}. Take a generic system operator $Y$, then 
\beq
\label{eq:qreg}
\begin{aligned}
\aver{ {Y (t)Y^\dagger (t') }  } &=  \operatorname{tr}\left[Y  \hat{\mathcal{V}}\left(t- t^{\prime}\right) Y^\dagger \rho \left(t^{\prime}\right)\right]  \theta(t-t') \\ 
&+ \operatorname{tr}\left\lbrace Y^\dagger  \hat{\mathcal{V}}\left(t^{\prime}- t \right) \lsq \rho \left(t \right) Y \rsq \right\rbrace  \theta(t'-t)
\end{aligned}
\eeq
Note that $ \aver{ {Y(t)Y^\dagger(t') }  } = \aver{ {Y(t')Y^\dagger(t) }  }  ^*  $ relates the $t>t'$ with the $t<t'$ expression. 
We remark that a similar result is not expected to hold for generic non-Markovian approaches and it is a peculiar property of NCA.
Note that the numerical cost for evaluating these correlation functions in NCA is the same as for computing $\hat{\mathcal{V}}$, making correlation functions easily accessible. An application to the spin-boson model is given in Sec. \ref{app:corrFunc_sb}.

\subsection{Validity and generalizations} 
% \subsection{Validity of NCA and benchmarking with higher-order self-consistent schemes} 
\label{sec:beyondNCA}

The NCA maps can be generalized to a whole hierarchy of self-consistent approximations, where the dissipators always depend on $\mcV$ rather than on $ e^{\mcH_s \tau }$. This is achieved using standard results of many-body theory that allow to rewrite self-energies in terms of \textit{skeleton} diagrams \cite{stefanucci2013nonequilibrium}. 
Each approximation in the hierarchy has contributions up to all powers in the system-bath coupling in the dissipator, still the hierarchy is ordered, such that higher-order terms become negligible for a sufficiently small system-bath coupling.
The NCA corresponds to lowest order approximation of the hierarchy. 

The NCA maps therefore become quantitatively correct when the coupling with the environment is sufficiently weak, such that the higher-order self-consistent corrections become negligible. 
We remark that they might still be quantitatively accurate in regimes in which Born-based master equations already show significant deviations from the correct solution, as we will demonstrate for the spin-boson model in Sec. \ref{sec:ocaSB}. 
%Finally, we recall that at lowest-order in the system-bath coupling, the NCA maps become equivalent to the Born master equation (see Sec. \ref{sec:fromBornToNCA}). 
Given their non-perturbative nature, the NCA maps might also be able to capture qualitatively the physics in the strong-coupling regime, as we show in Sec. \ref{sec:spinBoson}, while quantitative accuracy is not guaranteed there.

The natural strategy to assess the validity of the results of the NCA maps is to evaluate the leading-order self-consistent correction. This is known as ``one-crossing approximation'' (OCA) (see e.g. \cite{gullMillis2010,ecksteinWerner2010a}) and still has a compact expression adding one term to the NCA dissipator, that we derive and report in Appendix \ref{app:OCA}. We compute and discuss the OCA corrections for the spin-boson model in Sec. \ref{sec:ocaSB}. 
Note also that going one order beyond OCA is feasible \cite{ecksteinWerner2010a,kimEckstein2023} and a Monte-Carlo sampling around NCA is also possible \cite{gullMillis2010,cohen2014green}.

Compared to master equations, we remark that the decay time $\tau_b$ of the bath correlation function $\Gamma(\tau)$ is the shortest timescale for system-bath interactions in those equations \cite{lidarWhaley2001,majenzLidar2013,
breuerPetruccione2007,whitneyWhitney2008,
davidovicDavidovic2020}, while the NCA maps capture processes that happen on faster scales.
Also, for both NCA and Born approximations, 
$\tau_b$ is the decay time of the dissipator and therefore the timescale over which Eq. \eqref{eq:dyson} loses its ``memory'' of the past. 
Upon including higher-order self-consistent diagrams, the decay-time of the dissipator and thus the memory allowed by these approximations systematically increases; this is also shown in Fig. \ref{fig:seDiag}. Also remark that the NCA maps exactly preserve trace and Hermiticity of the density matrix \cite{schiroScarlatella2019}.

While so far we focused on a single bath of bosonic variables coupled to the system with their real part, 
the NCA maps generalize to multiple baths, either Markovian or treated at the same level of approximation, and extend to complex couplings such as in the rotating-wave approximation\cite{schiroScarlatella2019}, for which we report the dissipator in Appendix \ref{sec:rwa}.  
Also, these maps and their higher-order counterparts naturally generalize to fermionic
% , keeping track of their anti-commutation leading to different signs in the final expressions (the same as for Born-Markov master equations for the NCA maps); 
% They also generalize 
environments \cite{schiroScarlatella2019}, to non-stationary environments \cite{scarlatellaSchiro2021b} and to driven systems (time-dependent Hamiltonians).  
Instead, their derivation does not extend to spin environments, for which the Wick's theorem does not apply. Nevertheless, they might provide good approximations also in this case, for weak couplings and low temperatures such that there's few excitations in the environment that effectively behave like bosons.

\subsection{Numerical implementation and cost} 
\label{sec:numericalCost}
With respect to standard master equations based on the Born approximation \eqref{eq:BornEq}, equation \eqref{eq:dyson} for the dynamical map is a non-linear integro-differential equation. 
Non-linear equations can in principle lead to numerical instabilities, as for example one might encounter discretizing Eq.~\eqref{eq:dyson} with the simplest, explicit Euler scheme.
On the other hand, Eq.~\eqref{eq:dyson} has the form of a Dyson equation with a self-consistent self-energy, which is often encountered in many-body physics and for which several stable discretization schemes are known (see e.g.  \cite{aokiWerner2014 } and references therein). 
Here we use a simple and efficient implicit second-order Runge-Kutta scheme adapted from Ref. \cite{aokiWerner2014 } and described in Appendix \ref{app:rkScheme}, which is found to be numerically very stable (also in the case of a larger Hilbert space \cite{scarlatellaSchiro2021b}). 

Compared with Born-approximation-based master equations, the NCA maps 
can capture non-perturbative effects in the system-bath coupling, with a similar formalism and with a moderate increase in numerical cost. 
% The key advantage of the NCA maps over standard master equations based on the Born approximation is that they can capture stronger coupling regimes with a little extra numerical cost and with a similar formalism. 
The NCA map has a computational cost for time-propagation which is $O(t^2)$, like the Born master equation, and the NCA-Markov map has cost $O(t)$, as Redfield \cite{redfieldRedfield1957} or Lindblad equations \cite{breuerPetruccione2007}, allowing long propagation times. 
In the case of OCA~\eqref{eq:OCA} the cost gets an additional $O(t^2)$ factor, owing to the additional time-integrals involved. Note also that in the case of non-stationary environments or time-dependent Hamiltonians, there's an additional $O(t)$ cost, as the maps depend on two times rather than only on time differences.
The main computational disadvantage with respect to master equations is dealing with $\mcV$ which has size $N^4$, where $N$ is the size of the system Hilbert space, instead of with $\rho$ having size $N^2$.
This is not an issue for systems with a small Hilbert space, while for larger systems the additional cost might be limiting, but is still moderate compared to exact methods for open quantum systems. In the latter case of large systems, the maps can be used in combination with other many-body approaches such as the Dynamical Mean-Field Theory~\cite{georgesRozenberg1996,
aokiWerner2014,scarlatellaSchiro2021b} or could be compressed using memory-efficient methods such as matrix product operators~\cite{verstraeteCirac2004,zwolakVidal2004,flanniganDaley2021}. 

% While this is certainly not an issue for systems with a small Hilbert space, for larger systems these approaches might still be much cheaper than exact non-perturbative methods, and therefore might also be advantageous in combination with already numerically demanding many-body approaches \cite{scarlatellaSchiro2021b,flanniganDaley2021}.
%The NCA maps are therefore definitely advantageous 
%This is certainly not an issue for systems with a small Hilbert space, for which the NCA maps are certainly competitive with standard master equations. 

%The practical advantage of the NCA-Markov map is that it has a computational cost for the time-propagation which is $O(t)$, as standard Markovian master equations such as Redfield or Lindblad (the NCA map has cost $O(t^2)$ instead, like the Born master equation). 
%The practical advantage of the NCA-Markov map is that it has a computational cost for the time-propagation comparable to standard Markovian master equations, such as Redfield or Lindblad master equations.

\section{Application to the spin-boson model}
\label{sec:spinBoson}

In the following, we demonstrate the NCA and NCA-Markov dynamical maps on the spin-boson model and compare them to the Born and Born-Markov master equations.

This model  describes a quantum spin $1/2$ with tunneling strength $\Delta$ coupled to a bosonic bath with Hamiltonian
\begin{equation}
H=\frac{\Delta}{2} \sigma^{x}+\frac{\lambda}{2} \sigma^{z} \sum_{i}\left(a_{i}^{\dagger}+a_{i}\right)+\sum_{i} \omega_{i} a_{i}^{\dagger} a_{i}
\end{equation}
%The spin-boson is a prototype model of open quantum systems.
Despite its simple Hamiltonian, the spin boson model has a non-trivial physics that has been previously investigated with many theoretical methods~\cite{blumeLuther1970,brayMoore1982,chakravartyChakravarty1982,chakravartyLeggett1984,leggettZwerger1987,keilSchoeller2001,
andersSchiller2006,andersVojta2007,bullaVojta2005,vojtaVojta2012,amicoRibeiro2007,koppHur2007,leeZhang2011,
chinPlenio2011,beraFlorens2014,divincenzoLoss2005,luZheng2007,florensNarayanan2011,pineiroorioliRey2017,lindnerSchoeller2018,
yangTong2021}. It can also be nowadays realized in several experimental setups \cite{puertasmartinezRoch2019a,recatiZoller2005,magazzuGrifoni2018}. 
Identifying $X = \sigma^z  $ and $B = \frac{\lambda}{2}\sum_i \left( a^\da_i + a_i \right)$, then the bath correlation function $\Gamma(t)$ \eqref{eq:bathCorr}, is fixed by specifying the density of states of the bath modes and its temperature.
We consider here a zero temperature bath with bath correlation function
\beq
\Gamma(\tau) = \int_0^{\infty} \frac{d\xi}{2 \pi } J(\xi) \oh \lp \cos(\xi t) - i \sin(\xi t) \rp
\eeq
and the bath density of states is taken of the form \\ $J(\w) = \sum_i  \pi \lambda^2 \delta(\w - \w_i) = 2 \pi \alpha \w_c^{1-s}\w^s \theta(\w)\theta(\w_c-\w)$, with a sharp cutoff $\w_c$ and where $\alpha$ is the dimensionless system-bath coupling strength $\alpha = \lambda^2 / \lp 2 \w_c^2\rp $ (we fix $\w_c=1$ and $\Delta / \w_c =0.1$ throughout the manuscript).
For $s=1$ this corresponds to a Ohmic environment and for $s<1$ to a sub-Ohmic one, that will be considered in the rest of the manuscript.
\begin{figure*}[t]
\centering
\includegraphics[width=\linewidth]{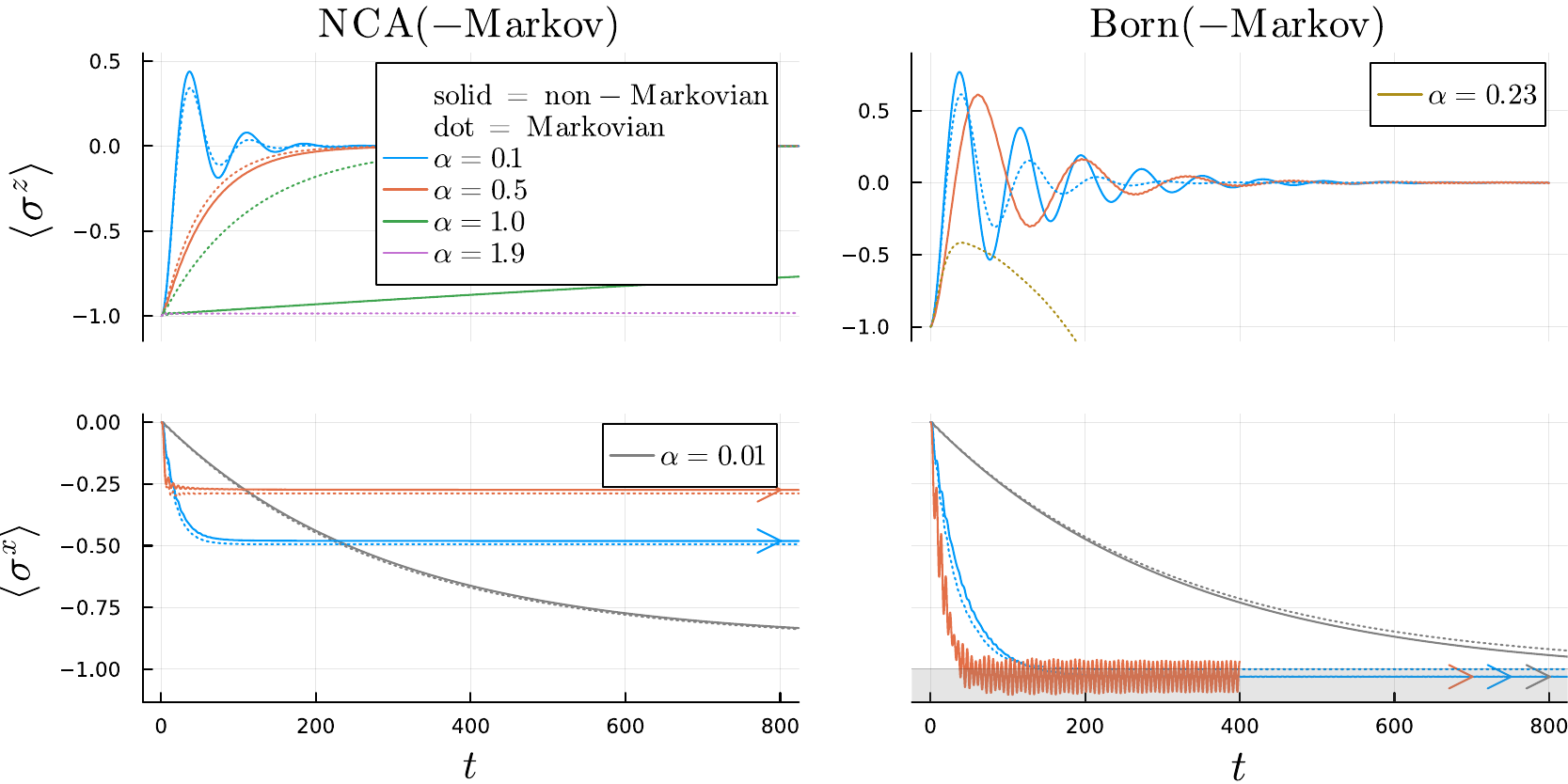}
\caption{Dynamics of the spin boson model, for $\Delta / \w_c =0.1, \w_c=1$, computed with the NCA and NCA-Markov maps (left) and with the Born and Born-Markov master equations (right) for different values of the system-bath coupling $\alpha$, starting from $\rho(0) = \ket{\downarrow} \bra{\downarrow  }$. 
The top panels show the dynamics of $\aver{\sigma^z}$: on the left, NCA(-Markov) captures that the dynamics becomes incoherent for $\alpha>\alpha_{\rm incoh}$; also, it captures that the timescale for relaxation diverges as $\alpha$ approaches a critical value $\alpha_c$, signaling the onset of the spin-boson delocalization transition. 
These features are not captured instead by the Born(-Markov) master equation (top-right). \\ 
The bottom panels show the relaxation of $\aver{\sigma^x}$ to its stationary value (indicated with arrows).
NCA(-Markov) predicts a non-trivial dependence of the steady-state on the system-bath coupling, which is missing in Born and Born-Markov, even at small coupling. 
The right panels show the breakdown of the Born approximation, developing unphysical oscillations (bottom), and of the Born-Markov one, developing an unphysical instability (top). 
%\new{We note that the latter instability can't be be associated to the development of a critical behaviour, as a finite relaxation timescale at short times is still present, as opposed to NCA. Also no signature of criticality is present in the Born approximation. }
Parameters: $dt=0.1 \times 2\pi/\w_c$, apart for the Born dynamics where $dt=0.01 \times 2\pi/\w_c$. 
% to mitigate the tendency to relax to unphysical states with $\aver{\sigma^x} < -1$ (see top-left panel). 
 }
\label{fig:dyn}
\end{figure*}

\subsection{Ohmic spin-boson: NCA maps vs master equations}
% \subsection{Ohmic spin-boson at strong coupling}

We consider here the Ohmic spin boson ($s=1$) in the limit $\Delta \ll \w_c$, that is well understood already from the seminal work~\cite{chakravartyLeggett1984}: for $\alpha < 1/2$, there are coherent oscillations (tunneling) between the $\sigma^z$ eigenstates at a renormalized spin frequency $\Delta_r$ which are damped by the environment, for $\alpha > 1/2$ then $\Delta_r$ is still finite but oscillations are overdamped and the dynamics is incoherent, while for $\alpha > \alpha_c =1 $ tunneling is suppressed and the spin enters a localized phase with $\Delta_r =0$, through a quantum-phase transition.

Fig. \ref{fig:dyn} shows the transient dynamics of the system starting from $\rho(0) = \ket{\downarrow} \bra{\downarrow  }$, with $\sigma^z \ket{\downarrow}  = -1  \ket{\downarrow}  $, with left and right columns corresponding to NCA maps and Born-based master equations. 
The spin-boson physics discussed above is captured both by the NCA and NCA-Markov maps, as shown in the 
top-left panel, plotting the dynamics of $\aver{\sigma_z}$. They capture a crossover between a coherent spin dynamics for $\alpha< \alpha_{\rm incoh}$, characterized by underdamped oscillations, and an incoherent dynamics for $\alpha > \alpha_{\rm incoh} $ where oscillations are overdamped.
Numerically we locate the crossover at $\alpha_{\rm incoh}\sim 0.2$, with a quantitative discrepancy with respect to more accurate estimates~\cite{chakravartyLeggett1984}.
We also note that in this regime the NCA and NCA-Markov agree remarkably well and that memory effects do not seem to play a major role.
Increasing the system-bath coupling, we see that both NCA and NCA-Markov approaches predict a growth of the spin relaxation timescale as $\alpha$ approaches the critical value, which witnesses the onset of the localization quantum phase transition \cite{andersSchiller2006}. While the qualitative behaviour of the two theories is similar, the NCA approach predicts $\alpha_c \approx 1$, as expected \cite{chakravartyLeggett1984}, while NCA-Markov predicts a larger value $\alpha_c \approx 1.9$.
The renormalization of the spin frequency down to zero at the localization transition can be equally inferred from the stationary-state correlation functions that we report in Appendix \ref{app:corrFunc_sb}.
Remarkably, the NCA and NCA-Markov maps can qualitatively capture the non-perturbative physics of the spin-boson model. 

On the other hand, the Born-based master equations fail to reproduce, even qualitatively, the crossover to an incoherent regime and the localization transition characterizing the spin boson model (see top-right panel).
The Born-Markov approximation is numerically divergent already at small coupling (top-right panel), while the Born one, despite not diverging, never predicts a crossover to over-damped oscillations (bottom-right panel).
That the Born-based master equations cannot predict the localization transition is also evident from the analysis of steady-state correlation functions in Appendix \ref{app:corrFunc_sb}.

The lower panels show instead the relaxation of $\aver{\sigma_x}$ to its stationary value (indicated by arrows), for different values of $\alpha$.   
For a vanishingly small system-bath coupling the spin thermalizes to its ground state, where $\aver{\sigma^x} = -1$, while increasing $\alpha$ the coupling with the bath drives the spin towards the $z$ direction, and thus $\aver{\sigma^x}$ decreases.
We remark that the NCA maps reproduce the expected behaviour in the whole delocalized phase (bottom-left panel). 
Instead, the Born and Born-Markov master equations cannot capture any steady-state dependence on $\alpha$ (see Appendix \ref{app:steadyState}), and thus they cannot predict the $\aver{\sigma^x}$ dynamics correctly (bottom-right panel). Similar pathologies are known to severely limit these equations, even at small couplings \cite{flemingCummings2011,tupkaryPurkayastha2022}. 
We also note that the Born master equation develops unphysical $\aver{\sigma_x}$ oscillations (bottom-right panel), reminiscent of an unphysical gap in the equilibrium $\sigma_x$ correlation functions \cite{florensNarayanan2011}, and unphysical $\aver{\sigma_x}<-1$ steady-state values of this quantity. 
These correspond to a non-positive density matrix, that instead the NCA maps never showed a tendency to develop, in the entire delocalized phase. 
% We remark here that instead the NCA maps never showed such a tendency. 
% to develop non-positive density matrices. %, while the Born master equation did (see top-right panel).
Finally, we remark that the NCA maps in the present form cannot capture the localized phase of the spin-boson model, in which the bath develops finite coherence and finite anomalous correlation functions, but they can be extended also to that case.
% Finally, for $\alpha$ larger than the critical value, we find (not shown) that both the NCA and NCA-Markov approaches become unstable, yielding density matrices with negative probabilities.
% Nota that the instability of the NCA approaches, though, only arises in correspondence of the localization transition, likely because the NCA approximation cannot capture the spontaneous development of finite coherences in the bath $\aver{B} \neq 0$ in the localized phase.

% and were also found to be numerically more stable then perturbative approaches.

In addition, to further highlight the ability of the NCA map to go beyond standard master equations, in Appendix \ref{app:corrFunc_sb} we computed the steady-state correlation functions showing how the spin transition frequency is renormalized to zero at the localization transition in NCA, while this never happens for Born-based master equations. 
In Appendix \ref{app:exp} we studied the response of the spin-boson model with a finite bias and shown that it reproduces the crossover to an incoherent regime observed experimentally in~\cite{magazzuGrifoni2018} that Born-based master equations fail at predicting. 

%we evaluated the response of the spin-boson model in presence of a finite bias in Appendix \ref{app:exp} and shown that it reproduces the qualitative features recently observed experimentally~\cite{magazzuGrifoni2018} 

\subsection{A quantitative benchmark}
% \subsection{Ohmic and sub-Ohmic dynamics at weak coupling}
\label{sec:ocaSB}

Here we show that the NCA maps can give results that are quantitatively accurate beyond the capability of Born-approximation-based master equations. 
We consider a spin-boson model with both Ohmic and sub-Ohmic ($s<1$) environments. In the latter case in fact, the localization transition happens at weaker system-bath couplings \cite{florensNarayanan2011}, influencing more the weak coupling regime where the NCA becomes quantitatively accurate. 

\begin{figure}[h]
\centering
\includegraphics[width=1\linewidth]{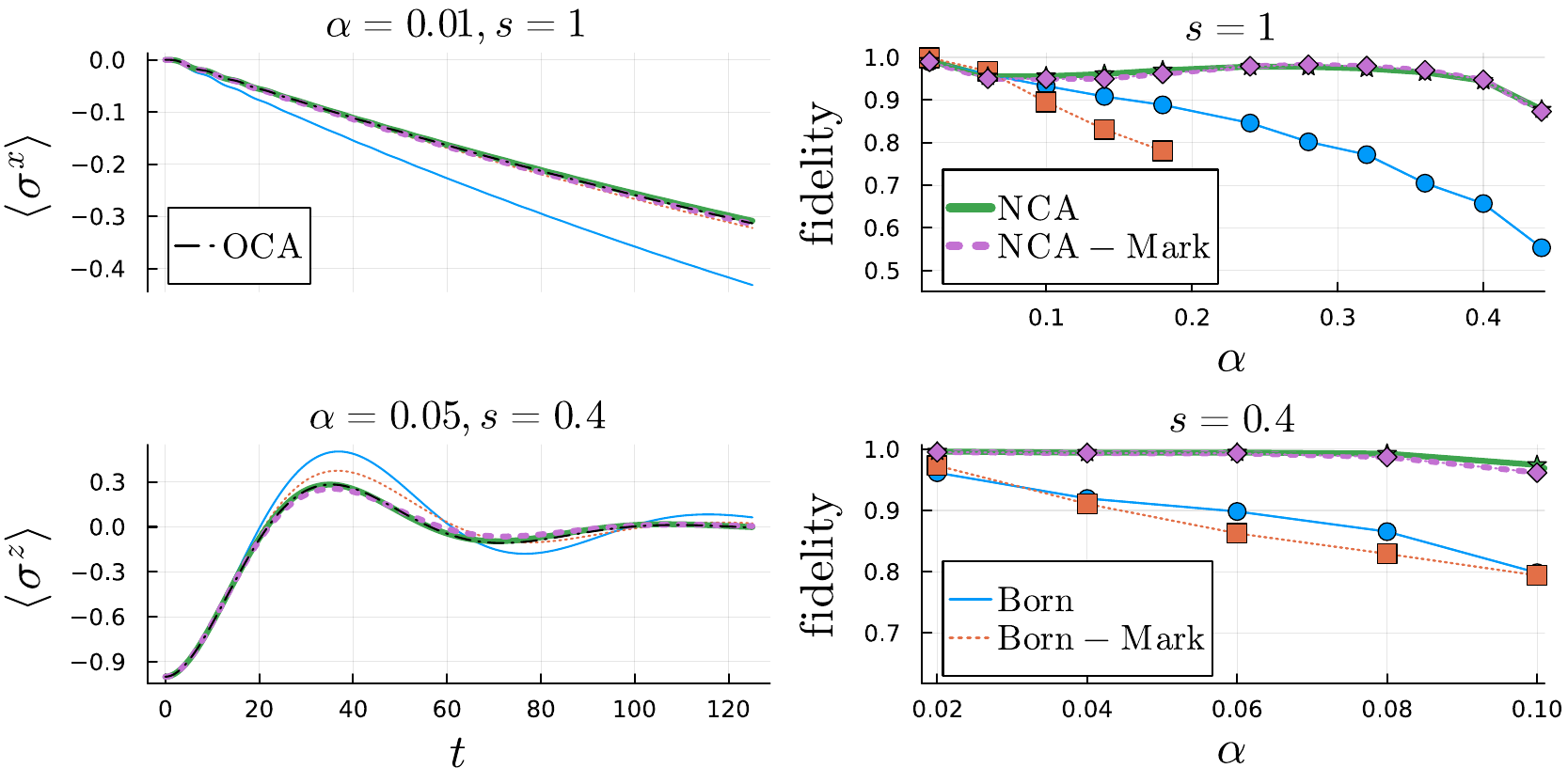}
\caption{A comparison between NCA, OCA and the Born approximation, for a Ohmic (top panels) and sub-Ohmic $s=0.4$ baths (bottom panels). The NCA predictions are shown to be quantitatively correct, since the OCA predictions are very close to the NCA ones. 
Respectively, the left panels show the real time dynamics, while left ones show the minimum process fidelity of the various maps compared to OCA as defined in the text. Note that in the upper-right plot, the Born-Markov points are plotted until $\alpha \approx 0.2$ as it diverges for larger couplings. The levels splitting and bath cutoff are $\Delta / \w_c =0.1, \w_c=1$.
}
\label{fig:ocaDyn}
\end{figure}

We compute the leading-order OCA corrections, that are compared in Fig. \ref{fig:ocaDyn} against the results of the NCA maps and of the Born-based master equations.
The left panels show the real-time dynamics, for which it can be seen that the OCA corrections to NCA are negligible at all times, while the Born and Born-Markov equations show significant deviations. 
The top-left panel shows the dynamics of $\aver{\sigma_x}$ for a Ohmic environment, for which we already remarked (in Fig. \ref{fig:dyn}) that the NCA maps capture a non-trivial steady-steady value, that the Born-based equations cannot predict. Accordingly, the latter equations display a large discrepancy also in the transient dynamics, while NCA results are accurate, as the OCA corrections are small. 
A similar conclusion is drawn from the bottom panel, where we show the dynamics of $\aver{\sigma_z}$ for a sub-Ohmic environment with $s=0.4$. In this case, the zero steady-state value is correctly captured by all methods, 
% For this quantity, all the methods yield the same correct steady state value, 
yet the master equations show significant errors in the transient dynamics. 

In the right panels instead we report an initial-state and observable-independent measure of the fidelity between the most accurate OCA map, and the other methods.  
As such a measure, we consider the Jamiolkowski process fidelity $F(\rho, \sigma) \equiv \operatorname{tr}(\sqrt{\sqrt{\rho} \sigma \sqrt{\rho}})^2$ of the density-matrices associated with the respective maps, that indeed describe quantum processes, by the Jamiolkowski isomorphism \cite{gilchristNielsen2005}:  $
\rho_{\mathcal{V}} \equiv[\mathcal{I} \otimes \mathcal{V}](|\Phi\rangle\langle\Phi|)
$, where $|\Phi\rangle=\Sigma_j|j\rangle|j\rangle / \sqrt{d}$ is a maximally entangled state of the ($d$-dimensional) system with another copy of itself, $\{|j \rangle \}$  is an orthonormal basis set and  $\mathcal{I}$ the identity matrix.
In the right panels of Fig. \ref{fig:ocaDyn} we show the fidelity with OCA, defined as the minimum fidelity in the time-interval identified by the time-axis of the left panels (a measure of worst error).  
For both the Ohmic and sub-Ohmic case the NCA maps have better values of fidelity, up to much higher values of coupling: for the Ohmic case the fedelity starts to deteriorate around $\alpha \approx 0.5$, after which the NCA (and OCA) results are still qualitatively but not quantitatively correct, while in the sub-Ohmic case the fidelity is good until $\alpha \approx 0.1$, that is close to the localization transition critical point \cite{florensNarayanan2011}.

We finally remark that the NCA-Markov approach in Fig. \ref{fig:ocaDyn} is in good agreement with NCA and OCA predictions, showing that the Born approximation is much more limiting than the Markovian approximation here.

\section{Conclusion}

%In this work we discussed how the Born approximation underlying standard weak-coupling master equations, such as Redfield and Lindblad equations, can be promoted to a self-consistent Born approximation, that is equivalent to a non-crossing approximation (NCA) in the system-bath coupling perturbation theory. 
%This yields an equation for the dynamical map of the system, the NCA map, that can also be combined with the usual Markovian approximation. 
%%In this work we have discussed how the NCA and NCA-Markov maps dynamical maps for open quantum systems. 
%We have shown that these approaches are formally very similar to conventional master equations based on the Born and Born-Markov approximation, and become equivalent to them in the weak-coupling limit, but can also capture stronger-coupling effects at very little extra numerical cost, and therefore might be preferred in several numerical studies of open quantum systems.

In this work we have introduced the NCA and NCA-Markov maps dynamical maps for open quantum systems. These approaches are formally
very similar to conventional master equations based on
the Born and Born-Markov approximation and become
equivalent to them in the weak-coupling limit, but can
also capture non-perturbative effects and stronger couplings with a moderate increase in numerical cost and therefore might be preferred in several numerical studies of open quantum systems.
We have also discussed how the NCA maps can be benchmarked evaluating the leading order self-consistent correction, the one-crossing approximation (OCA). 

%go beyond them in capturing strong-coupling effects in the system-bath coupling which occur on a shorter timescale than the typical decay time of bath correlation functions. 
We applied the NCA maps to the spin-boson model at zero temperature, showing how they qualitatively capture its non-perturbative physics in presence of a Ohmic bath, including its crossover between coherent and incoherent dynamics and its phase transition towards a localized phase, signalled by a growth of the spin relaxation time-scale and a renormalization of the spin frequency to zero, which conventional weak-coupling master equations fail at predicting.  
We have also evaluated the leading-order OCA correction both for a Ohmic and a sub-Ohmic spin-boson model, showing that the NCA maps also yield quantitatively correct predictions in regimes in which standard master equations show significant deviations. 
%discussed how the NCA predictions can be tested introducing its leading order OCA correction. We showed for the Ohmic and sub-Ohmic spin-boson model that the NCA maps can yield both qualitative and also quantitatively correct predictions, in regimes in which standard master equations cannot.

%Apart from capturing qualitatively new physics at a comparable numerical cost than perturbative master equations, the NCA and NCA-Markov maps were found to be 
%well behaved for much stronger system-bath couplings, owing to their highly non-linear nature. Therefore, these techniques might be preferred to the conventional master equations in numerical studies of open quantum systems.
%
%We expect the NCA maps to find a variety of applications beyond the model considered in this paper. Among these we mention the experimentally-relevant driven spin-boson model in the regime of coherent dynamics~ \cite{magazzuGrifoni2018}, systems of quantum emitters coupled to nanophotonic structures \cite{gonzalez-tudelaCirac2017,gonzalez-tudelaCirac2017a,gonzalez-tudelaCirac2018} or electronic transport problems ~\cite{espositoGalperin2009,espositoGalperin2010,jinYan2014}. Furthermore they can be used in connection with Dynamical Mean-Field Theory of open quantum many-body systems~\cite{scarlatellaSchiro2021b}, as impurity solvers for lattice problems or to explore strong system-bath coupling effects on extended quantum systems~ \cite{flanniganDaley2021,maierRoos2019}.

We expect the NCA maps to find a variety of applications beyond the model considered in this paper, including electronic transport problems 
% where standard master equations are not sufficient to account for lifetime broadening effects 
~\cite{espositoGalperin2009,espositoGalperin2010,jinYan2014}, driven problems~\cite{yamaguchiOgawa2017,dimeglioHuelga2023,aharonovichToth2016}, non-perturbative environments such as nanophotonic structures~\cite{gonzalez-tudelaCirac2017,gonzalez-tudelaCirac2017a,gonzalez-tudelaCirac2018} and collective effects of multiple emitters coupled to a common environment \cite{sheremetPoddubny2023}.

In the case of systems with a large Hilbert space, these maps can be combined with a Dynamical Mean-Field Theory approach~\cite{georgesRozenberg1996,aokiWerner2014,scarlatellaSchiro2021b}, or could be compressed using memory-efficient methods such as matrix product operators~\cite{verstraeteCirac2004,zwolakVidal2004,flanniganDaley2021}.

\section*{Acknowledgements}

\paragraph{Funding information}
We acknowledge computational resources on the Coll\'ege de France IPH cluster. This work was supported by the ANR grant ”NonEQuMat” (ANR-19-CE47-0001).
It was also supported by the Engineering and Physical Sciences Research Council (EPSRC) and by the Science and Technology Facilities Council (STFC) [grant number EP/W005484].

%\bibliography{}
%\end{document}

\pagebreak

\appendix

\section{Implementation of a 2nd order Runge-Kutta scheme}
\label{app:rkScheme}

Equation \eqref{eq:dyson} is a Volterra integral-differential equation, that often appears in many-body problems, with the form 
\begin{equation}
\label{eq:volt}
\frac{d}{d t} y(t)=q(t)+p(t) y(t)+\int_0^t d \bar{t} k(t, \bar{t}) y(\bar{t})
\end{equation}
where in our case of an equilibrium bath the memory kernel depends only on time differences.  
Several stable discretization schemes can be used to solve it and here we implemented a 2nd order Runge-Kutta scheme adapted from \cite{aokiWerner2014 }, Appendix A, to propagate it in time. 

In the present case we adapt that scheme, such as to 
capture correctly the case in which the memory kernel has a $\delta(t)$ contribution, for example in the case of a Markovian environment. 
By integrating \eqref{eq:volt} up to the time-steps $t_n$ and $t_{n-1}$ and subtracting the two equations, we get
\begin{equation}
y\left(t_n\right)-y\left(t_{n-1}\right)=\Delta t \sum_{i=0}^{n-1}\left(w_{n, i}-w_{n-1, i}\right) y^{\prime}\left(t_i\right)+\Delta t w_{n, n} y^{\prime}\left(t_n\right)
\end{equation}
which is Eq. A5 of \cite{aokiWerner2014 }, with $w_{n,i}$ ($i=(0,1, \dots, n)$) some weights that depend on the scheme chosen for approximating integrals (Euler, trapezoids, Simpson's rule, ecc). 
$y^{\prime}\left(t_n\right)$ is then evaluated by discretizing \eqref{eq:volt}. In doing so, we allow $k(t,\bar{t})$ to have a $\delta(t-\bar{t})$ contribution. 
To correctly take it into account, we discretize the integral as 
\begin{equation}
\es{
\int_0^{t_n} d \bar{t} k\left(t_n, \bar{t}\right) y(\bar{t}) &= \int_0^{t_{n-1}} d \bar{t} k\left(t_n, \bar{t}\right) y(\bar{t}) + \int^{t_n}_{t_{n-1}} d \bar{t} k\left(t_n, \bar{t}\right) y(\bar{t}) = \\
&= \Delta t \sum_{i=0}^{n-1} w_{n-1, i} k\left(t_n, t_i\right) y\left(t_i\right) + \Delta t k\left(t_n, t_n\right) y(t_n)
}
\end{equation}
where we discretized the last time-step separately, to make sure that $k\left(t_n, t_n\right)$ is evaluated with weight 1, such as to fully capture a $\delta(t-\bar{t})$ contribution (instead of discretizing the whole integral with a single scheme).

In our implementation, we use a second-order Runge-Kutta scheme for the weights $w_{n,i}$, corresponding to 
\begin{equation}
w_{n, i}= \begin{cases}1 / 2 & i=0, n \\ 1 & 1 \leq i \leq n-1\end{cases}
\end{equation}
leading to 
\begin{equation}
\es{
y^{\prime}\left(t_n\right) & = q\left(t_n\right)+p\left(t_n\right) y\left(t_n\right)+ \Delta t   k\left(t_n, t_n\right)y\left(t_n\right) + \\
&\frac{\Delta t}{2} \lsq k\left(t_n, t_0\right)y\left(t_0\right) + k\left(t_n, t_{n-1}\right) y\left(t_{n-1}\right)  \rsq 
+ \Delta t \sum_{i=1}^{n-2} k\left(t_n, t_i\right) y\left(t_i\right)
}
\end{equation}

\begin{equation}
y\left(t_n\right)-y\left(t_{n-1}\right)=\frac{\Delta t}{2} \lsq y^{\prime}\left(t_{n-1}\right)  +y^{\prime}\left(t_n\right) \rsq
\end{equation}

Using those equations it is possibile to determine $y(t_n),y^\prime (t_n)$, given their values for $t_{n'}<t_{n}$ and an initial condition for $y(t_0),y^\prime (t_0)$ (in our case $\mcV(0) = \id$ and $\pt \mcV(0)$ is given by Eq. \eqref{eq:dyson}). 
If $k\left(t, \bar{t}\right)$  doesn't depend on $y$ one might eliminate $y^\prime(t_n)$ and find a closed equation for $y(t_n)$. 
Since in our case $k\left(t, \bar{t}\right)$ depends on $y$ as well, we iterate the 2 equations above to find $y(t_n),y^\prime (t_n)$.

\section{System-bath coupling in the Rotating Wave Approximation}
\label{sec:rwa}

Here we consider the case of a system-bath coupling in the form $\tilde{H}_{SB} = \tilde{X} \otimes \tilde{B}^\da + \tilde{X}^\da \otimes \tilde{B}$, which is common when performing a rotating-wave approximation. Here $\tilde{B}= \sum_i \frac{ \tilde{\lambda}_i}{2} a_i $. We still consider a bosonic bath, while for fermionic ones we refer to \cite{schiroScarlatella2019}.  In this case the NCA dissipator is given by 
\begin{equation}
\begin{aligned}
\hat{\tilde{\mathcal{D}}}(\tau)= & \tilde{\Gamma}^{<} (\tau) \left(\tilde{X}^{\da} \mcV(\tau)[\bullet ] \tilde{X}- \mcV(\tau)[\bullet \tilde{X}] \tilde{X}^{\da}\right)+ \\
& \tilde{\Gamma}^{>} (\tau) \left(\mcV (\tau)[\tilde{X} \bullet] \tilde{X}^{\da}-\tilde{X}^{\da} \mcV (\tau)[\tilde{X} \bullet ]\right)+ \hc
\end{aligned}
\end{equation}
with $\tilde{\Gamma}^{>}(\tau) = \tr \lsq \tilde{B}(\tau) \tilde{B}^\da (0) \rho_B(0) \rsq $ and $\tilde{\Gamma}^{<}(\tau) = \tr \lsq \tilde{B}^\da(\tau) \tilde{B} (0) \rho_B(0) \rsq $.

%In the case of a system-bath coupling of the rotating-wave-approximation form in \ref{sec:rwa}, 
% In this case, the correlation functions take the form 
% \begin{equation}
% \begin{aligned}
% \aver{ \tilde{X} (t) \tilde{X}^{\da} (t') }  = &  \operatorname{tr}\left[ \tilde{X} \hat{\mathcal{V}}\left(t, t^{\prime}\right) \tilde{X}^{\dagger} \rho\left(t^{\prime}\right)\right]\theta(t-t') \\ &+  \operatorname{tr}\left\lbrace \tilde{X}^\da \hat{\mathcal{V}}\left(t^\prime, t\right) \lsq \rho\left(t \right) \tilde{X}  \rsq \right\rbrace\theta(t'-t) \\ 
% \aver{ \tilde{X}^\da (t) \tilde{X} (t') }  = &  \operatorname{tr}\left[ \tilde{X}^{\dagger} \hat{\mathcal{V}}\left(t, t^{\prime}\right) \tilde{X} \rho\left(t^{\prime}\right)\right]]\theta(t-t')  \\ 
% & +  \operatorname{tr}\left\lbrace \tilde{X} \hat{\mathcal{V}}\left(t^{\prime}, t\right) \lsq \rho\left(t\right) \tilde{X}^\da \rsq\right\rbrace ]\theta(t'-t)
% \end{aligned}
% \end{equation} 

\section{Formal derivation of the self-consistent dynamical maps}
\label{app:formalDer}

\subsection{Perturbation theory in the system-bath coupling}

In this section we derive the perturbation series for the evolution superoperator in the system-bath coupling, up to all orders in this coupling. 
This leads to the Dyson equation \eqref{eq:dyson}
of the main text and sets the stage for developing the non-crossing approximation.

We recall that the evolution superoperator $\mcV$ is defined by $\rho(t) = \tr_B \rho_{\rm tot}(t)  = \mcV(t) \rho(0)$ where $\rho_{\rm{tot}}(t) = e^{-i H t} \rho_{\rm{tot}}(0) e^{i H t}$
and the total Hamiltonian is given by $H= H_S+ H_B + H_{SB}$ and where $\tr_B$ is a partial trace on bath operators. 
By moving to the interaction picture, the following identities can be found for the evolution operators
\begin{align}
e^{- i H t} = e^{- i \lp H_S + H_B \rp t } T_t e^{- i \int_0^t dt' H_{SB}(t')} = e^{- i \lp H_S + H_B \rp t } T_t \sum_{n =0}^\infty \frac{(-i)^n}{n!} \lsq \int_0^t dt' H_{SB}(t') \rsq^n  \\ 
e^{i H t} = \check{T}_t e^{ i \int_0^t dt' H_{SB}(t')} e^{ i \lp H_S + H_B \rp t } =  \check{T}_t \sum_{n=0}^\infty \frac{(+i)^n}{n!} \lsq \int_0^t dt' H_{SB}(t') \rsq^n e^{ i \lp H_S + H_B \rp t }
\end{align}
Here $H_{SB}(t') = e^{i (H_S+H_B) t'} H_{SB} e^{-i (H_S + H_B) t'}$ and $T_t$ is the real-time time-ordering operator. The latter takes any product of operators, where each operator is defined at one time, and changes the order so that every operator has only later operators to the left and earlier operators to the right. For non-fermionic variables as those we will consider, this change of order doesn't introduce any sign. The anti-time-ordering operator $\check{T}_t$ instead enforces the opposite ordering.
Plugging in these expressions and using the cyclic property of the partial trace on the bath, one finds
\small
\begin{align}
\label{eq:hyb_befSup}
 \mcV(t) \rho(0)  & =   e^{-i H_S t} \tr_B \lbr T_t \sum_{{k_+}=0}^\infty \frac{(-i)^{k_+}}{{k_+}!} \lsq \int_0^t dt' H_{SB}(t') \rsq^{k_+}  \rho_{\rm{tot}}(0) \check{T}_t \sum_{{k_-}=0}^\infty \frac{(+i)^{k_-}}{{k_-}!} \lsq \int_0^t dt' H_{SB}(t') \rsq^{k_-}  \rbr e^{i H_S t} 
\end{align}
\normalsize

In order to manage this double series expansion, it is convenient to use the superoperators notation already introduced in the main text. We define
\beq
\hat{H}_{SB \gamma}(t') \bullet  = 
\begin{cases}
H_{SB}(t') \bullet & \rm{if}\quad \gamma= +  \\ 
\bullet H_{SB}(t')  & \rm{if}\quad \gamma= -
\end{cases}
\eeq
Using this definition, we note that the time-ordering structure of the operators in Eq. \eqref{eq:hyb_befSup}, that are time-ordered on the left of the density matrix and anti-time-ordered on its right, is obtained in the superoperator notation if superoperators with $\gamma=+$ and $\gamma=-$ are time-ordered separately, putting superoperators with later times on the left regardless of their index $\gamma$ (instead of those with $\gamma=-$ being anti-time-ordered). We therefore introduce the operator $T_C$ enforcing this time-ordering of a string of superoperators (the notation recalls that of the contour-time-ordering operator in Keldysh field theory, which is analogous).
This is easily understood with an example. Suppose $t_1>t_2>t_3>t_4$, then
\beq
\es{
T_C \hat{O}_-(t_3) \hat{O}_-(t_4) \hat{O}_+(t_1) \hat{O}_+(t_2) \bullet &= \hat{O}_-(t_3) \hat{O}_-(t_4) \hat{O}_+(t_1) \hat{O}_+(t_2)  \bullet = \\ = O(t_1) O(t_2) \bullet O(t_4) O(t_3) &=
T_t O(t_1) O(t_2) \bullet \check{T}_t O(t_3) O(t_4)
}
\eeq
We also note that superoperators with different $\gamma$ indexes commute by definition, \\ $\hat{O}_+ (t_1) \hat{O}_- (t_2)  \bullet = \hat{O}_-(t_2)\hat{O}_+(t_1) \bullet =O(t_1) \bullet O(t_2)  $, thus we can always assume that a string of superoperators under $T_C$ is time-ordered according to their real-time variable, regardless of their indexes $\gamma$.
Using the previous example
\beq
\es{
T_C \hat{O}_-(t_3) \hat{O}_-(t_4) \hat{O}_+(t_1) \hat{O}_+(t_2)  &= \hat{O}_-(t_3) \hat{O}_-(t_4) \hat{O}_+(t_1) \hat{O}_+(t_2) = \\ &= \hat{O}_+(t_1) \hat{O}_+(t_2) \hat{O}_-(t_3) \hat{O}_-(t_4) 
}
\eeq
where in the last equality the superoperators are ordered according to their real time variables $t_1>t_2>t_3>t_4$.
We also define the bare evolution superoperator $\mcV_0(t) = e^{-i H_S t} \bullet e^{i H_S t} $. 
With these definitions one can write
\begin{align}
 \mcV(t) \rho(0) & =   \mcV_0(t) \tr_B \lbr T_C \sum_{{k_+},{k_-}=0}^\infty \frac{(-i)^{{k_+}+{k_-}}}{{k_+}!{k_-}!} \lsq \int_0^t dt' \hat{H}_{SB+}(t') \rsq^{k_+}  \lsq - \int_0^t dt' \hat{H}_{SB-}(t') \rsq^{k_-}   \rho_{\rm{tot}}(0) \rbr 
\end{align}
By doing some combinatorial calculations, it's easy to show that the double summation can be reduced to a single one and that the following identity is valid:
\begin{align}
 \mcV(t) \rho(0)  & =  \mcV_0(t) \tr_B \lbr T_C  e^{- i \int_0^t dt' \sum_{\gamma \in \{+,-\}} \gamma \hat{H}_{SB\gamma}(t') } \rho_{\rm{tot}}(0) \rbr = 
\\
 &= \mcV_0(t) \tr_B \lbr T_C \sum_{k=0}^\infty \frac{(-i)^{k}}{k!} \lsq  \int_0^t dt' \sum_{\gamma \in \{+,-\}} \gamma \hat{H}_{SB\gamma}(t')   \rsq^k \rho_{\rm{tot}}(0) \rbr 
\end{align}

We assume that there are initially no correlations between the system and the bath, that is $\rho_{\rm{tot}} (0)= \rho(0) \otimes \rho_B$, such that we can perform the partial trace over bath operators stemming from $H_{SB}(t') = X(t') \otimes B(t') $, raised to the $k$-th power. 
We take the bath state to be stationary under the bare bath Hamiltonian, for example to be in equilibrium at some bath temperature, and, without of loss of generality, we assume that $\tr_B{\lp B(t) \rho_B(0)\rp}  = \tr_B{\lp B \rho_B(0)\rp} = 0$: a finite value of $\tr_B{\lp B \rho_B(0)\rp} $ could be in fact always absorbed in the definition of the Hamiltonian \cite{nathanRudner2020}. Under this assumption, only terms with even powers of $k$ are non-zero, and therefore we change the summation index to sum only on those even terms.
Finally we consider system and bath operators that are non-fermionic, such that they commute under time-ordering. Under these assumptions, we obtain
%Assuming that there are initially no correlations between the system and the bath, thus $\rho_{\rm{tot}} = \rho(0) \otimes \rho_B$, and that system and bath operators are non-fermionic, such that they commute under time-ordering, and recalling that $H_{SB}(t') = X(t') \otimes B(t') $, we obtain

\beq
\es{
 &\mcV(t) \rho(0) = \mcV_0(t)    \sum_{k=0}^\infty \frac{(-i)^{2k}}{(2k)!} \int_0^t dt_1 \dots \int_0^t dt_{2k} \times \\
 &\sum_{\gamma_1 \dots \gamma_{2k}} \gamma_1 \dots \gamma_{2k}\tr_B \lsq T_C \hat{B}_{\gamma_1}(t_1) \dots \hat{B}_{\gamma_{2k}}(t_{2k}) \rho_B(0) \rsq T_C \hat{X}_{\gamma_1}(t_1) \dots \hat{X}_{\gamma_{2k}}(t_{2k}) \rho(0)
 }
\eeq

Since the bath Hamiltonian $H_B = \sum_i \w_i a^\da_i a_i$ is quadratic in its creation and annihilation operators and assuming also its initial state $\rho_B(0)$ is, we can use the Wick's theorem to simplify the multi-point correlators of bath operators.
For Hermitian bosonic operators such as  $B = \sum_i \frac{\lambda_i}{2} \lp a_i + a_i^\da \rp $ one can use the Wick's theorem for real bosonic variables yielding
\beq
\tr_B \lsq T_C \hat{B}_{\gamma_1}(t_1) \dots \hat{B}_{\gamma_{2k}}(t_{2k}) \rho_B(0) \rsq = 
\sum_{ \underset{\lbr (\gamma_1,t_1),\dots (\gamma_{2k},t_{2k}) \rbr }{ \mathrm{pairings\,of }}}   \Gamma_{\gamma_{i_1} \gamma_{i_2}} (t_{i_1} - t_{i_2} ) \dots \Gamma_{\gamma_{i_{2k-1}} \gamma_{i_{2k}}} (t_{i_{2k-1}} - t_{i_{2k}} )   
\eeq
On the left-hand side, the two-point correlation functions of the bath appear
\beq 
\label{eq:time_ord_2_point}
\Gamma_{\gamma ,\gamma ' }(t-t') = \tr_B{ \lsq T_C \hat{B}_\gamma(t)\hat{B}_{\gamma ' }(t') \rho_B(0) \rsq} ,
\eeq 
which depend on time differences because we assumed $\rho_B(0)$ to be stationary, and the sum runs over all the possible ways of forming pairs with the indexes $\lbr (\gamma_1,t_1),\dots (\gamma_{2k},t_{2k}) \rbr$. 
Also, since the integrand is completely symmetric by permuting two integration/summation indexes $(t_i,\gamma_i)$ and $(t_j,\gamma_j)$, we can limit the integration to the domain defined by $t_1>t_2>\dots >t_{2k}$ and multiply by $(2k)!$. 
We can also drop the contour time-ordering operator $T_C$, which is at this point useless since time-ordering is already enforced by the extremes of integration.
Finally, we write the string of system operators by making explicit their time-evolution as follows
\beq
\mcV_0(t) \hat{X}_{\gamma_1}(t_1) \dots \hat{X}_{\gamma_{2k}}(t_{2k}) = \mcV_0(t-t_1) \hat{X}_{\gamma_1}  \mcV_0(t_1-t_2) \dots \hat{X}_{\gamma_{2k}}\mcV_0(t_{2k}) 
\eeq
and we drop the initial state $\rho(0)$, since this is arbitrary.
Eventually, we obtain the series for the evolution superoperator $\mcV$: 
\beq
\label{eq:hybrExp}
\es{
&\mcV(t)  = \sum_{k=0}^\infty (-i)^{2k} \int_0^{t} dt_1 \int_0^{t_1} dt_2 \, \dots  \int_0^{t_{2k-1}} dt_{2k} \sum_{ {\gamma_1 \dots \gamma_{2k} } } \gamma_1 \dots \gamma_{2k} \sum_{ \underset{\lbr (\gamma_1,t_1),\dots (\gamma_{2k},t_{2k}) \rbr }{ \mathrm{pairings\,of }}} \\ 
& \lsq  \Gamma_{\gamma_{i_1} \gamma_{i_2}} (t_{i_1} - t_{i_2} ) \dots \Gamma_{\gamma_{i_{2k-1}} \gamma_{i_{2k}}} (t_{i_{2k-1}} - t_{i_{2k}} ) \rsq  
\mcV_0(t-t_1) \hat{X}_{ \gamma_1} \mcV_0(t_1-t_2) \hat{X}_{ \gamma_2}  \dots \hat{X}_{\gamma_{2k}}\mcV_0(t_{2k}) ,
}
\eeq

While it is possible to manipulate this series directly, it is more convenient for our purposes to show that it takes the form of the Dyson equation \eqref{eq:dyson} of the main text and to introduce its self-energy. This is the object of the next section.

\subsection{Feynman diagrams, self-energy and Dyson equation}
%
%Expanding in the system-bath interaction and after some manipulations similar to Ref. \cite{schiroScarlatella2019}, one obtains the following series expansion for the evolution superoperator $\mcV$:  \cmm{can I have V not dep on time-diff?}

%Here $\mcV_0(t - t') \bullet = e^{i H_S (t-t')} \bullet e^{-i H_S (t-t')} $ and $\gamma_i\in \{+,-\}$. The last sum runs over all the possible ways of forming pairs with the indexes $\lbr (\gamma_1,t_1),\dots (\gamma_{2k},t_{2k}) \rbr$ and results from applying the Wick's theorem for real bosonic variables \cite{kamenevKamenev2011,altland_simons_2010}. Those pairs appear as arguments of the two-point correlation functions of the bath 
%\beq 
%\label{eq:time_ord_2_point}
%\Gamma_{\gamma ,\gamma ' }(t-t') = \tr{ \lsq T_C B_\gamma(t)B_{\gamma ' }(t') \rho_B(0) \rsq}, 
%\eeq 
%
%These are time-ordered correlation functions on the Keldysh contour, meaning that the contour time-ordering operator $T_C$ acts as follows 
%%orders the operators $B_\gamma(t), B_{\gamma ' }(t')$ according to 
%\begin{align}
%%(t,\gamma) > (t' ,{\gamma '} ) 
%T_C B_\gamma(t) B_{\gamma ' }(t') &= B_\gamma(t) B_{\gamma ' }(t')
% &\text{if} 
%\begin{cases}
%t>t'  &\gamma = \gamma ' =+  \\ 
%t<t' &\gamma = \gamma ' = -  \\ 
% \gamma = - &\gamma ' = +
%\end{cases} \nonumber\\
%T_C B_\gamma(t) B_{\gamma ' }(t') &=  B_{\gamma ' }(t')B_\gamma(t)  &\rm{otherwise} \nonumber
%\end{align}
In order to show that Eq. \eqref{eq:hybrExp} can be cast in the Dyson form of Eq. \eqref{eq:dyson} of the main text, we start by introducing its diagrammatic representation.
Each term of the series with fixed $k, t_1\dots t_{2k}, \gamma_1 \dots \gamma_{2k}$ and with a fixed choice of pairings can be represented as a Feynman diagram following these rules: 
\begin{itemize}
\item draw two parallel solid lines, each representing a portion of time axis going from time $t=0$ to time $t$.
\item locate the $t_1 > \dots > t_{2k}$ times on those axes, where the $k_+$ times corresponding to an $\hat{X}_+$ superoperator must be drawn on the first time axis and the $k_-=k-k_+$ times corresponding to an $\hat{X}_-$ superoperator on the other one. The couple of solid-line segments going from $t_i$ to $t_{i+1}$ represent the evolution superoperator $\mcV_0(t_{i+1}-t_i)$. 
%draw $2k$ times on those contours corresponding to the times at which the operators $X$ are introduced, respectively $n$ on the $+$(upper) and $m$ on the $-$(lower) axis, with $n+m=2k$. Keep track of the factor $(-1)^{2k}$.
\item connect the times paired by $\Gamma$ functions with dashed lines.
\end{itemize}
Finally, one also needs to keep track of the factor $(-1)^{2k}$ as well as of the sign given by the product $\gamma_1 \dots \gamma_{2k}$. 
As an example, the term with expression 
\beq
\label{eq:sample_diagm}
\Gamma_{+ -}(t_1-t_2) \mcV_0 (t-t_1) \hat{X}_{+} \mcV_0(t_1-t_2)  \hat{X}_{-} \mcV_0(t_2) \bullet  
\eeq
corresponds to the diagram in Fig. \ref{fig:sample_diag}:
\begin{figure}[H]
\center 
\includegraphics[width=0.17\linewidth]{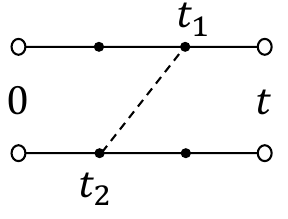}
\caption{The diagrammatic representation of the term \eqref{eq:sample_diagm}.} \label{fig:sample_diag}
\end{figure}
Summing over the $\gamma$ indexes, one can also define more compact diagrams where the double-time axis is collapsed on a single time-axis, as shown in Fig. \ref{fig:compact_diag}:
%The sub-series of diagrams obtained by summing over the $\gamma$ indexes, corresponding to the terms in which each operator appears either on the $+$ or on the $-$ contour, can be conveniently be represented by a simplified diagram, in which the double-time contour is collapsed on a single time axis, as shown in Fig. REF.
\begin{figure}[H]
\center
\includegraphics[width=0.9\linewidth]{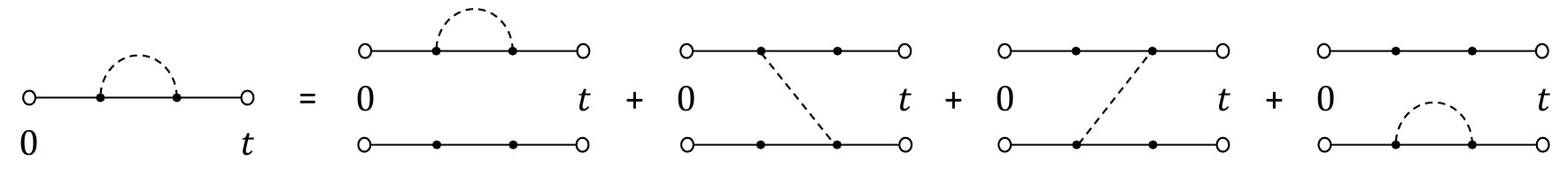}
\caption{Single time-axis diagrams represent the set of diagrams where super-operator indexes are summed over.}
\label{fig:compact_diag}
\end{figure}
From these diagrammatic rules, we can define a ``self-energy'' of the series on the lines of textbook calculations in many-body theory.
For this purpose, we define as one-particle-irreducible (1PI) the compact diagrams which cannot be separated, by cutting a solid line, in two parts that are not connected by any dashed line. An example of 1PI diagrams is given in Fig. \ref{fig:1PI_cross}. Then, the self-energy $\hat{\mathcal{D}}$ is defined as the sum of all 1PI diagrams with the first an last solid lines removed: its diagrammatic representation is given in Fig. \ref{fig:1PI_cross}. 

%\begin{figure}[h]
%\center
%\includegraphics[width=0.55\linewidth]{fig1_2}
%\caption{
%Diagrammatic representation of the self-energy: the solid lines correspond to time evolution superoperators $e^{\mcH_s (\tau)}$, while the dashed lines correspond to 2-times bath correlation functions. %(see \cite{Note1}).  
%The NCA consists in retaining only the diagrams in which the dashed lines do not cross. 
%Their sum $S_{\rm NCA}$ has a closed form, represented in terms of a bold solid line that stands for the dynamical map $\mcV$. 
%We remark that all the non-crossing diagrams decay on the timescale $\tau_b$ in which bath correlation functions decay. 
%}
%\label{fig:dyson_se}
%\end{figure}

\begin{figure}[H]
\center
\includegraphics[width=0.9\linewidth]{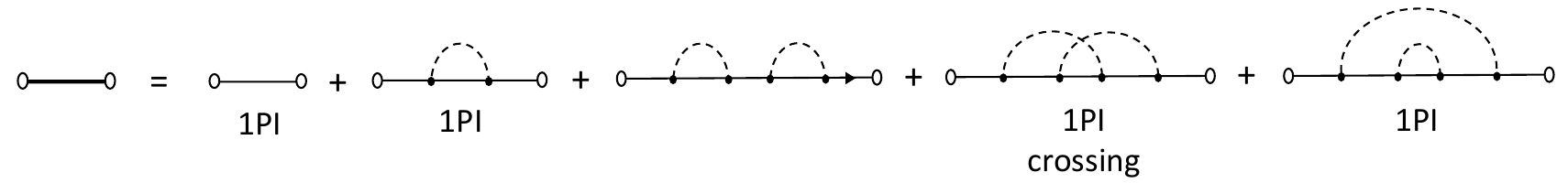}
\caption{1-particle-irriducible (1PI) diagrams making up the self-energy and crossing diagrams.}
\label{fig:1PI_cross}
\end{figure}

The superoperator $\mcV$ is given by the sum of all diagrams, both 1PI and non-1PI. The definition of the self-energy is useful because all non-1PI diagrams can be obtained by joining some 1PI diagrams with solid lines and, therefore, the whole series for $\mcV$ can be written as
$$\mcV  = \mcV_0 + \mcV_0 \circ \hat{\mathcal{D}} \circ \mcV_0 + \mcV_0 \circ \hat{\mathcal{D}} \circ \mcV_0 \circ \hat{\mathcal{D}} \circ \mcV_0  + \dots , $$ where the circle operator ``$\circ$'' stands for partial time convolutions as in Eq. \eqref{eq:hybrExp}.
This series sums up to the Dyson equation 
\beq
\es{
\mcV(t)  = \mcV_0(t) + \int_0^{t} dt_1 \int_0^{t_1} dt_2 \mcV_0(t-t_1) \hat{\mathcal{D}} (t_1-t_2) \mcV(t_2) 
\\ = \mcV_0(t) + \int_0^{t} dt_1 \int_0^{t_1} dt_2 \mcV(t-t_1) \hat{\mathcal{D}} (t_1-t_2) \mcV_0(t_2)
}
\eeq
or equivalently, to the integro-differential form reported as Eq. \eqref{eq:dyson} in the main text:
\beq
%\label{eq:dyson}
\partial_{t} \hat{\mathcal{V}}(t)=\hat{\mathcal{H}}_S \hat{\mathcal{V}}(t)+\int_{0}^{t} d t_{1} \hat{\mathcal{D}}\left(t-t_{1}\right) \hat{\mathcal{V}}\left(t_{1}\right)
\eeq

\subsection{The non-crossing approximation}

The non-crossing approximation (NCA) corresponds to approximating the series for $\mcV $, and thus also for $\hat{\mathcal{D}}$, by keeping only the compact diagrams in which dashed lines do not cross. 
We remark that considering the compact diagrams is important for defining non-crossing diagrams, as in their non-compact version ambiguities arise as dashed lines might cross or not depending on how one draws them.
An example of crossing diagram is given in Fig. \ref{fig:1PI_cross} and the diagrammatic representation of the NCA self-energy is shown in Fig.~\ref{fig:seDiag}.  

It turns out that the NCA self-energy coincides with the $k=1$ term of the exact self-energy, where the bare propagator $\mcV_0$ is replaced with the dressed one $\mcV $. This statement corresponds to the second equality in Fig.~\ref{fig:seDiag}. 
In fact, we remark that the first and last times of a self-energy diagram must be connected together by a dashed line in the non-crossing approximation: if it's not the case, in fact, the resulting diagram is either non-1PI or it's crossed. We also remark that all diagrams with first and last time connected are necessarily 1PI.
Therefore all the diagrams of $\hat{\mathcal{D}}_{\rm{NCA}}$ are obtained connecting the remaining times in all possible non-crossing ways, but the latter diagrams in turn sum up to $\mcV $, proving the equality.

The explicit expression of the NCA self-energy is then given by
\beq
\label{eq:self-en}
\hat{\mathcal{D}}_{\rm{NCA}}(\tau) = (-i)^2 \sum_{\gamma_1 \gamma_2} \gamma_1 \gamma_2 \Gamma_{\gamma_1 \gamma_2} (\tau) \hat{X}_{\gamma_1} \mcV(\tau) \hat{X}_{\gamma_2} . 
\eeq
Carrying out the sum over $\gamma$ indexes and after some algebraic manipulations to go from the time-ordered correlation function $\Gamma_{\gamma_1, \gamma_2}(\tau)$, defined in Eq. \eqref{eq:time_ord_2_point}, to the non-time ordered correlation function $\Gamma(\tau)$, defined in Eq. \eqref{eq:bathCorr} of the main text, the NCA self-energy reduces to Eq. \eqref{eq:ncaSe} of the main text.

%In fact, we remark that the first and last times of a non-crossing self-energy diagram must be connected together by a dashed line: if it's not the case, in fact, the resulting diagram is either non-1PI or it's crossed. We also remark that all diagrams with first and last time connected are necessarily 1PI. Therefore all the diagrams of $\hat{\mathcal{D}}_{\rm{NCA}}$ are obtained connecting the intermediate times to form all the possible non-crossing diagrams, but the latter diagrams in turn sum up to $\mcV $, proving the equality. The NCA self-energy then coincides with the $k=1$ term of the exact self-energy, where the bare propagator $\mcV_0$ is replaced with the dressed one $\mcV $: 

\subsection{Validity of the NCA and Markov approximations together}
\label{app:validityNCAMarkov}

We discuss in this appendix the validity of the NCA-Markov approximation, leading to Eq. \eqref{eq:markDys} of the main text. At first glance the assumptions behind the NCA and Markovian approximations seem to contradict each other. 
The Markovian approximation assumes that the bath-induced dynamics on the timescale of bath correlation functions, call it $\tau_b$, is negligible, thus one can approximate
$\mcV(t-\tau)$ with 
$e^{-\mcH_s \tau} \mcV(t)$.
The NCA self-energy, one the other hand, captures bath-induced processes happening on timescales shorter than $\tau_b$. 
This contradiction is resolved and the two approximations can be made simultaneously, if one assumes that bath-induced dynamics, on a timescale $\tau_b$, can be neglected only when $t>\tau_b$. In this way the Markovian approximation on the NCA master-equation is legitimate:
\small
\beq
\label{eqapp:markApp}
\es{
&\partial_{t} \hat{\mathcal{V}}(t) \\ & =\hat{\mathcal{H}}_S \hat{\mathcal{V}}(t)+\int_{0}^{t} d \tau  \lsq    \Gamma  (\tau)   \lp - X_{  +} +X_{  -} \rp \mcV(\tau) X_{  +}  + \hc \rsq  \hat{\mathcal{V}}\left( t - \tau \right) \\ 
&\approx \hat{\mathcal{H}}_S \hat{\mathcal{V}}(t)+\int_{0}^{t} d \tau  \lsq    \Gamma (\tau)   \lp - X_{  +} +X_{ -} \rp \mcV(\tau) X_{  +}  + \hc \rsq   e^{- \mcH_s \tau} \hat{\mathcal{V}} \left( t \right)
}
\eeq
\normalsize
where $ \hat{\mathcal{V}} \left( t \right) \approx  e^{ \mcH_s \tau} \hat{\mathcal{V}}\left( t - \tau \right) $ for $t>\tau_b$ has been approximate in the spirit of a Markovian approximation, while $\mcV(\tau)$ is probed inside the integral only at times $\tau < \tau_b$ and thus bath-induced dynamics is not neglected for this propagator. 
The essence of the NCA-Markov approximation, therefore, is that the bath-induced dynamics at short times $t<\tau_b$ is feedbacked into the dynamics at $t > \tau_b$. 
%This master equation is then expected to be valid in intermediate-coupling regimes, where the bath-induced dynamics at times $t>\tau_b$ is still slower than the decay of bath correlation functions, justifying a Markovian approximation, but when the bath strongly renormalizes the short-time dynamics. \cmm{is this the case in spin-bos propagator?}
To be fully consistent, one should evolve $\mcV(t)$ up to $t \sim  \tau_b$ with the non-Markovian NCA equations and then continue the propagation using the NCA-Markov master equation. The results of the main text have been obtained instead by integrating the NCA-Markov equations from $t=0$.

We remark that the dynamics on the timescale $\tau_b$ is left completely unresolved in most Markovian master equations \cite{lidarWhaley2001,whitneyWhitney2008,davidovicDavidovic2020}, where the upper integration integral in Eq. \eqref{eqapp:markApp} is sent to infinity. 

%\begin{figure}{h}
%\includegraphics[width=0.3\linewidth]{selfEnOCA.pdf}
%\caption{The diagrammatic representation of the OCA self-energy}
%\label{fig:ocaSE}
%\end{figure}

\subsection{Beyond NCA}
\label{app:OCA}

%\new{\textit{Beyond the NCA maps--}
%The whole self-energy can be written in terms of a series of ``skeleton'' diagrams, which depend on $\mcV$ rather than on $ e^{\mcH_s \tau }$ \cite{stefanucci2013nonequilibrium}. 
%Each diagram has contributions up to all powers in the system-bath coupling, still the series is ``ordered'', such that higher-order diagrams become smaller than lower-order ones for a sufficiently small system-bath coupling.
%%This series is ``ordered'' such that higher-order diagrams become smaller than lower-order ones, when the system-bath coupling is sufficiently small.
%Truncating the series to different orders yields a family of self-consistent dynamical maps, of which the NCA corresponds to the first order. 
%Taking into account higher-order diagrams gives a natural strategy to assess the validity of the results of the NCA maps. Adding the leading-order correction, shown in Fig. \ref{fig:seDiag} and whose expression is reported in \cite{Note1}, is known as ``one-crossing approximation'' (OCA) (see e.g. \cite{gullMillis2010,ecksteinWerner2010a}).
%We also remark that, upon including higher-order diagrams of the skeleton series, the decay-time of the self-energy or ``memory'' of the bath systematically increases, as Fig. \ref{fig:seDiag} shows for the first two terms. 
%More details are given in \cite{Note1}.}

It is possible to go systematically beyond the NCA, which is particularly useful to assess the validity of its predictions. 
The exact self-energy $\hat{\mathcal{D}}$, which is represented in Fig. \ref{fig:seDiag} as a series of diagrams in terms of the ``bare'' propagator $\mcV_0 = e^{\mcH_s \tau }$, can be also expressed as a series of ``skeleton'' (dressed) diagrams, which depend only on $\mcV$ rather than on $\mcV_0$.
This is a standard result of many-body theory (see e.g. \cite{stefanucci2013nonequilibrium}). 
This skeleton series is defined by all the diagrams which have no self-energy insertions, that is no pieces that disconnect from the diagram by cutting two solid lines \cite{stefanucci2013nonequilibrium}.
%In this skeleton series, the diagrams are ordered in terms of the number of crossings of the lines representing 2-times bath correlation functions (dashed lines in our diagrams).  
The NCA yields the first term of the skeleton series for the self-energy. 
We remark that each diagram of the skeleton series contains contributions up to all powers in the system-bath coupling. 
Nevertheless an ``order'' in the series exists such that higher-order diagrams are smaller than lower-order ones for a sufficiently weak coupling, as it is the case for the bare series. 
The natural strategy to assess the validity of the NCA results is therefore to include higher-order contributions to the self-energy. At sufficiently small coupling the NCA predictions become quantitatively accurate, as higher-order diagrams are negligible, while at stronger coupling one can still check qualitative agreement upon including higher-order terms.
Adding the leading-order diagram beyond NCA yields the self-energy 

\beq
\label{eq:OCA}
\es{
\hat{\mathcal{D}}_{\rm OCA} (\tau) = &\hat{\mathcal{D}}_{\rm NCA}(\tau) + (-i)^4 \sum_{\gamma_1\gamma_2 \gamma_0\gamma}  \gamma_1\gamma_2 \gamma_0\gamma \times \\ 
&\int_0^\tau d \tau_1 \int_0^{\tau_1} d \tau_2 \Gamma_{\gamma \gamma_2} (\tau -\tau_2) \Gamma_{\gamma_1 \gamma_0} (\tau_1) \hat{X}_\gamma \mcV(\tau-\tau_1) \hat{X}_{\gamma_1} \mcV(\tau_1-\tau_2) \hat{X}_{\gamma_2} \mcV(\tau_2) \hat{X}_{\gamma_0} 
}
\eeq

This is known as one-crossing approximation (OCA) \cite{ecksteinWerner2010a,gullMillis2010}, as it corresponds to summing the bare diagrams in which 2 dashed lines, corresponding to bath 2-times correlation functions, cross at most once. 
The second term of \eqref{eq:OCA} becomes smaller than the first at a small enough coupling, because it is involves two bath correlation functions.
%Considering ``higher-order'' diagrams in the skeleton series in fact, the decay-time of the self-energy or ``memory'' of the bath is systematically increased, as Fig. \ref{fig:seDiag} shows for the first two terms. 
%When the coupling is sufficiently small, the NCA predictions become quantitatively accurate as higher-order diagrams become negligible. At stronger coupling instead, one can still check qualitative agreement with OCA or higher-order schemes. 
As we mentioned in the main text, diagrams of the skeleton series are also ordered in terms of increasing decay-time of the self-energy or ``memory'' of the bath. 
In Fig. \ref{fig:seDiag} we show that the NCA self-energy decays in a timescale $\tau \sim \tau_b$, in which the bath correlation function $\Gamma(\tau)$ decays: this is the case because $\hat{\mathcal{D}}_{\rm NCA}(\tau)$ is proportional to $\Gamma(\tau)$ (Eq. \eqref{eq:ncaSe} of the main text).
The second term in Eq. \eqref{eq:OCA} instead decays in $2 \tau_b$, because of the two bath correlation functions ($\Gamma_{\gamma \gamma'}$ reduces to $\Gamma$ or to its complex conjugate given the time-ordering of its arguments).

An application of the OCA to the Spin-Boson model is reported in \ref{sec:ocaSB}

\section{The Born-Markov master equation}
\label{app:weakCoupME}

A Markovian approximation of the Born master equation \eqref{eq:BornEq} leads to 
\beq
\partial_{t} \rho(t)=\hat{\mathcal{H}}_S \rho(t)+\int_{0}^{t} d \tau \hat{\mathcal{D}}_{\rm{Born}} \left(\tau \right) e^{-\mcH_s \tau} \rho(t)
\label{eq:born-mark}
\eeq
which together with \eqref{eq:bornSe} is the Born-Markov master equation, used in the main text for comparison with the NCA-Markov map.

The most common way of writing this equation (see e.g. \cite{mozgunovLidar2020}) is by explicitly replacing the self-energy \eqref{eq:bornSe} in \eqref{eq:born-mark} getting
%to 
%\beq
%\label{eq:app1.2}
%\partial_{t} \rho(t)=\hat{\mathcal{H}}_S \rho(t)+\int_{0}^{t} d \tau   \lsq \Gamma (\tau)   \lp - \hX_{+} +\hX_{ -} \rp e^{ \mcH_s \tau }  X_{ +}  e^{ - \mcH_s \tau } + \hc \rsq \rho(t)
%\eeq
%which looks familiar. This can be written as 
\beq
\label{eq:app1.3_blochRed}
\partial_{t} \rho(t)= - i \lsq H_s, \rho(t) \rsq  + \lsq X \tilde{X} \rho(t) +  \tilde{X}  \rho(t) X + \hc  \rsq 
\eeq
in terms of the ``filtered'' operators  
\beq 
\label{eq:filt_op}
\tilde{X}   = \int_{0}^{t} d \tau \Gamma (\tau)    e^{-i H_s \tau} X e^{ i H_s \tau} 
\eeq
%and used the identity $e^{ \mcH_s \tau }  \hX_{ \gamma}  e^{ - \mcH_s \tau }  = \lp e^{ - i H_s \tau }  \hX  e^{ i H_s \tau } \rp_\gamma $. 

In the main text, we always keep the integration time in \eqref{eq:filt_op} finite to compare with the NCA-Markov map. 
Instead, when doing a fully Markovian approximation the integration in \eqref{eq:filt_op} is extended up to infinity \cite{whitneyWhitney2008}, as bath correlation functions decay on a timescale  $\tau \sim \tau_b$, while the system is assumed to change on a much slower timescale. 
This allows to evaluate the operator $\tilde{X} $ in the basis of the eigenstates of the system $\ket{E_n}$ with eigenvalues $E_n$, where it is expressed in terms of the Fourier transform of the retarded component of the bath correlation function $\Gamma^R (\w) = \int_{-\infty}^{\infty} d\tau e^{i \w \tau } \Gamma (\tau)  \theta(\tau)$:
\beq 
\tilde{X}   = \sum_{nm} \Gamma^R (E_m-E_n) \bra{n} X \ket{m} \ket{n} \bra{m}
\eeq

\section{Further results on the Spin-Boson model}

\subsection{Steady state correlation functions}
\label{app:corrFunc_sb}

In the main text, we have discussed the transient dynamics of the Ohmic spin-boson model, while here we focus on its correlation functions
computed once the steady-state of the dynamics has been reached.
%  thermal equilibrium between the spin and the bath has been reached as a steady-state of the dynamics. \cmm{doubt: can i use same prop in steady-state? also staeady-state for born is unphysical...}

%\textit{The quantum regression theorem} for Markovian master equations allows to compute multi-time correlation functions from the same generator of the dynamics which evolves the density matrix. 
%This result is generalized in NCA \cite{schiroScarlatella2019,scarlatellaSchiro2021b}, at least for the two-times correlation functions of the operator $X$ which couple to the bath, which can be computed from the formula 
%
%
%\beq
%\label{eq:qreg}
%\begin{aligned}
%\aver{ {X (t)X (t') }  } =  \operatorname{tr}\left[X  \hat{\mathcal{V}}\left(t, t^{\prime}\right) X \rho \left(t^{\prime}\right)\right]  
%\end{aligned}
%\eeq
%The formula is valid for $t>t'$, while for $t<t'$ one can use the relation $ \aver{ {X(t)X(t') }  } = \aver{ {X(t')X(t) }  }  ^*  $ and compute the correlation function in the right-hand side using Eq. \eqref{eq:qreg}.
%We remark that a similar result is not expected to hold for generic non-Markovian approaches and it is a peculiar property of NCA.
\begin{figure}
\center
\includegraphics[width=0.65\linewidth]{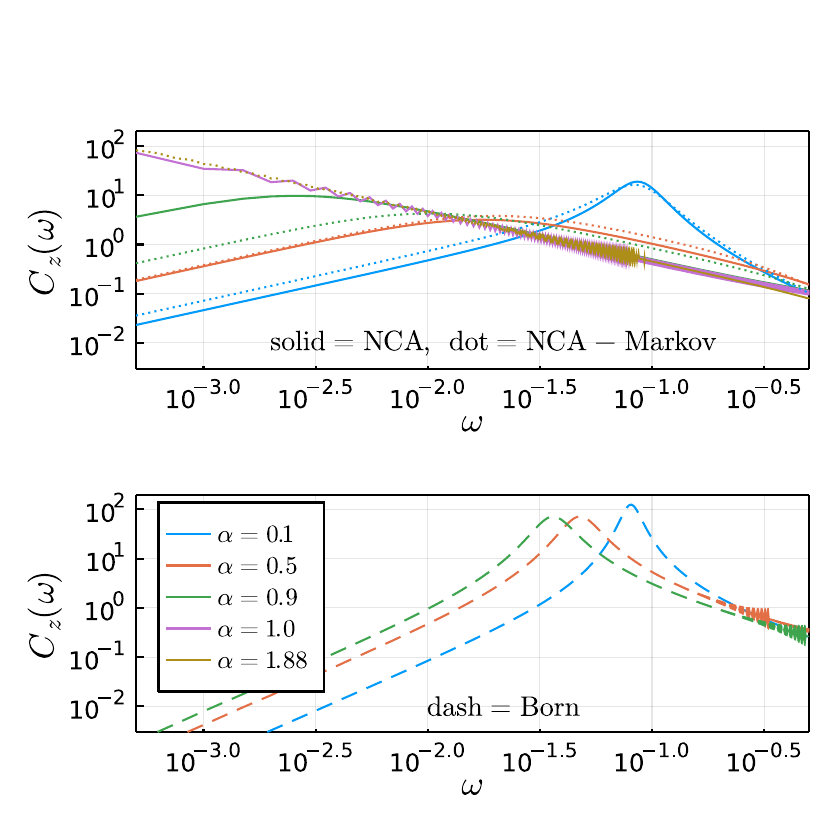}
\caption{Steady-state correlation function of the $z-$spin component, $C_z(\omega)$ obtained through the NCA and NCA-Markov (top panel)  quantum dynamical maps using Eq.~(\ref{eq:qreg}), and their Born approximation (bottom panel). We see that both NCA approaches correctly capture the renormalization of the spin frequency $\Delta_r$ due to the bath coupling, ultimately leading to a quantum phase transition into a localized phase when $\Delta_r =0$ for $\alpha=\alpha_c$. The low frequency behavior shows a power-law behavior of the form $C_z(\omega)\sim \omega$ as expected for an ohmic bath. On the other hand the Born approximation fails both in the capturing the frequency renormalization and in the power-law regime at low frequency, different from the expected linear behavior. }
\label{fig:corr}
\end{figure}

In Fig. \ref{fig:corr} we show the Fourier transform of the steady-state correlation function \\
$C_z(t) = \oh \lim_{t'\rw \infty} \aver{ \lsq \sigma_z(t+t'), \sigma_z(t') \rsq }$, which is a purely real function, computed from Eq. \eqref{eq:qreg}.
At weak coupling with the bath, this correlation function shows a peak at $\Delta$, corresponding to spin-flip transitions at the frequency of the bare spin. 
Increasing the coupling to the bath, the spin frequency gets renormalized to a smaller value $\Delta_r$, which appears as a shift of the corresponding peak in  $C_z(\w)$. 
The upper panel of Fig. \ref{fig:corr} shows that, increasing the system-bath coupling, both the NCA and NCA-Markov approaches predict such a renormalized spin frequency $\Delta_r$, which approaches $\Delta_r =0 $ as the critical coupling $\alpha_c$ is reached. This is known to happen \cite{leggettZwerger1987,chakravartyLeggett1984} for the Ohmic spin-boson model at zero temperature for $\Delta/\w_c \ll 1$. 
Note though that the exact functional dependence of $\Delta_r$ on $\alpha$ close to the critical point $\Delta_r=c \Delta(\Delta/\w_c)^{\alpha/(1-\alpha)}$ \cite{leggettZwerger1987,kehreinNeu1995} (where $c$ is a constant prefactor) is not correctly reproduced by the NCA. The Born-Markov approximation is not shown in Fig. \ref{fig:corr} as it is numerically unstable for $\alpha=0.5,0.9$ considered, while the Born approximation predicts only a very small shift of the peak, which never reaches zero for the values of coupling explored (up to $\alpha\approx 200$).
%On the other hand (bottom panel) the Born approximation predicts only a very small shift of the peak, which never reaches zero for the values of coupling explored ($\alpha\approx 200$). The Born-Markov approximation is not shown as it is unstable for $\alpha=0.5,0.9$ considered in Fig. \ref{fig:corr}, while predicting a resonance at a frequency $\w$ in agreement with the Born approximation for $\alpha=0.1$.

In addition, we also remark that the power-law behaviour of $C_z(\w)$ at small $\w$ is correctly captured in the NCA and NCA-Markov approximations, $C_z(\w) \sim \w $, while the Born approximation predicts $C_z(\w) \sim \w^2$.

\subsection{Connections to experiments}
\label{app:exp}

In a recent experiment \cite{magazzuGrifoni2018}, the spin-boson model is realized by a superconducting transmission line, whose electromagnetic modes represent the bosonic environment, coupled to a qubit, realizing the two-state system. 
The response of the system can be probed by applying a weak probe field at a frequency $\omega$ to the transmission line and measuring its transmission
\beq
\mathcal{T}(\omega) = 1 - i \mathcal{N} \omega \chi(\omega)
\eeq
where $  \chi(\omega) (t-t') = -i \aver{ \lsq \sigma^z(t), \sigma^z(t') \rsq} \theta(t)$ is the retarded Green function of $\sigma^z$ and $\mathcal{N}$ is a coupling constant that we set to $\mathcal{N}=1$, as we aim at a qualitative comparison with Ref. \cite{magazzuGrifoni2018}. 

The crossover between underdamped-coherent and incoherent dynamics of the spin boson model can probed by transmission measurements scanning both the frequency of the probe $\w$ and a static magnetic flux applied to the qubit $\epsilon$, resulting in the qubit Hamiltonian

\begin{equation}
H_{\rm qb}=\frac{\Delta}{2} \sigma^{x}+\frac{\epsilon}{2} \sigma^{z} 
\end{equation}

\begin{figure*}
\includegraphics[width=1\linewidth]{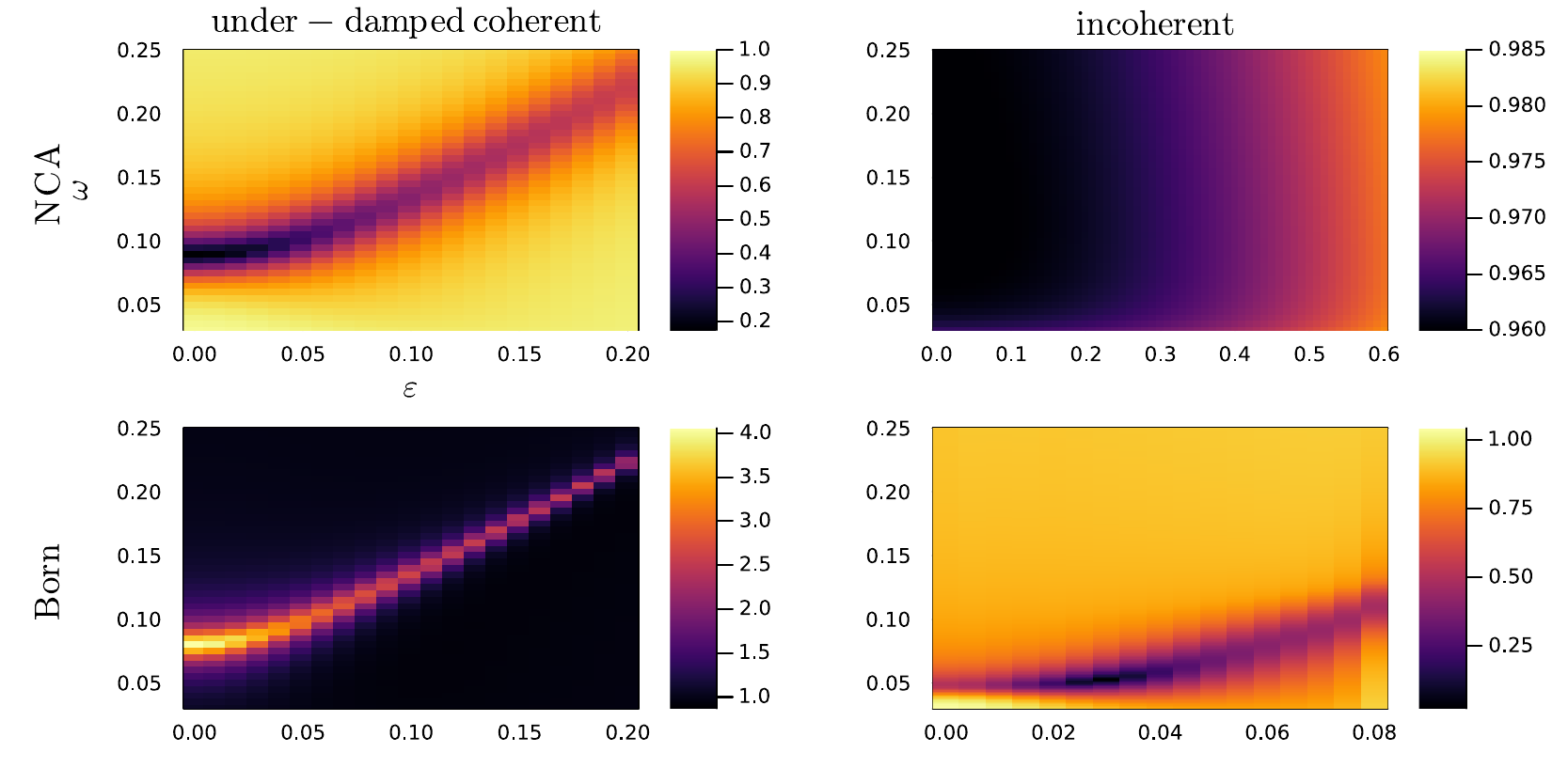}
\caption{Modulus squared transmission $\abs{\mathcal{T}(\omega)}^2$ of an applied probe field of frequency $\w$ as a function of the two-level system bias $\epsilon$. 
The top panels are obtained using the NCA map, while the bottom ones with a Born master equation. 
Left(right) panels correspond to the regime of underdamped-coherent (incoherent) dynamics for $\alpha=0.1(0.6)$.  
The NCA map captures the crossover between underdamped and incoherent dynamics, while the Born master equation always predict an underdamped dynamics.
}
\label{fig:transm_4pan}
\end{figure*}

In Fig. \ref{fig:transm_4pan} the color code indicates the modulus squared transmission $\abs{\mathcal{T}(\omega)}^2$ as a function of $\w$ and $\epsilon$ (in units of $\w_c$). 
The first column refers to the regime of underdamped-coherent dynamics for $\alpha =0.1$ while the second column to the incoherent dynamics regime for $\alpha=0.6$; the top panels are obtained using the NCA map, while the bottom ones with a Born master equation. 
The NCA results are in qualitative agreement with the experimental results of \cite{magazzuGrifoni2018} (their Fig. 2), showing that the qubit dispersion relation can be read out of the transmission measurement in the underdamped-coherent regime (left panel), while in the incoherent regime (right panel) the transmission is nearly independent of the probe frequency $\w$. 
In the bottom panels we show that the Born master equation does not capture the transition to a regime of incoherent dynamics and always predicts a trace of the qubit dispersion in the transmission (bottom right panel). This result agrees with the results on the real-time dynamics described in the main text (Fig. \ref{fig:dyn}). 
We also remark that the Born master equation wrongly predicts enhanced, rather than suppressed, transmission for values of $\w$ and $\epsilon$ hitting the qubit dispersion for $\alpha=0.1$ (left plot).

\subsection{Steady-state dependence in NCA vs Born}
\label{app:steadyState}

The steady-state equation \eqref{eq:steadyState} also allows to understand why the NCA approaches capture the dependence of the steady-state on the system-bath coupling $\alpha$ in the top panels of Fig. \ref{fig:dyn} of the main text, which is instead missing in Born and Born-Markov theories: from our calculations we find that the steady-state of the spin-boson model does not depend on the Hamiltonian and takes the form $\rho_s =  p_- \ket{-} \bra{-} +p_+ \ket{+} \bra{+}$. In this case then Eq. \eqref{eq:steadyState} reduces to $\int_{0}^{\infty} d \tau \hat{\mathcal{D}}(\tau) \rho_{s}=0$. 

By expanding the self-energy $\hat{\mathcal{D}}(\tau)$ in powers of the system-bath coupling $\alpha$ as \\
$\hat{\mathcal{D}}(\tau)  =\alpha \hat{\mathcal{D}}^{(1)}(\tau) +  \alpha^2 \hat{\mathcal{D}}^{(2)}(\tau) + \dots  $, the Born and Born-Markov approximations correspond to truncating at first order in $\alpha $ and thus $\alpha$ drops in the steady-state equation: $ \int_{0}^{\infty} d \tau \hat{\mathcal{D}}^{(1)}(\tau) \rho_{s}=0$. 
Instead, the NCA and NCA-Markov approaches retain higher-order contributions to the self-energy, yielding a non-trivial dependence of the steady-state on $\alpha$.

%\bibliography{}
%\end{document}
\bibliography{./paper.bib}

\begin{thebibliography}{100}
\providecommand{\url}[1]{\texttt{#1}}
\providecommand{\urlprefix}{URL }
\expandafter\ifx\csname urlstyle\endcsname\relax
  \providecommand{\doi}[1]{doi:\discretionary{}{}{}#1}\else
  \providecommand{\doi}{doi:\discretionary{}{}{}\begingroup
  \urlstyle{rm}\Url}\fi
\providecommand{\eprint}[2][]{\url{#2}}

\bibitem{wangsnessBloch1953}
R.~K. Wangsness and F.~Bloch,
\newblock \emph{The {{Dynamical Theory}} of {{Nuclear Induction}}},
\newblock Phys. Rev. \textbf{89}(4), 728 (1953),
\newblock \doi{10.1103/PhysRev.89.728}.

\bibitem{redfieldRedfield1957}
A.~G. Redfield,
\newblock \emph{On the {{Theory}} of {{Relaxation Processes}}},
\newblock IBM Journal of Research and Development \textbf{1}(1), 19 (1957),
\newblock \doi{10.1147/rd.11.0019}.

\bibitem{kuboKubo1969}
R.~Kubo,
\newblock \emph{{A Stochastic Theory of Line Shape}},
\newblock In \emph{{Advances in Chemical Physics}}, pp. 101--127. {John Wiley
  \& Sons, Ltd},
\newblock ISBN 978-0-470-14360-5,
\newblock \doi{10.1002/9780470143605.ch6} (1969).

\bibitem{scullyLamb1967}
M.~O. Scully and W.~E. Lamb,
\newblock \emph{Quantum {{Theory}} of an {{Optical Maser}}. {{I}}. {{General
  Theory}}},
\newblock Phys. Rev. \textbf{159}(2), 208 (1967),
\newblock \doi{10.1103/PhysRev.159.208}.

\bibitem{mollowMiller1969}
B.~R. Mollow and M.~M. Miller,
\newblock \emph{The damped driven two-level atom},
\newblock Annals of Physics \textbf{52}(3), 464 (1969),
\newblock \doi{10.1016/0003-4916(69)90289-9}.

\bibitem{mccauleyJacobs2020}
G.~McCauley, B.~Cruikshank, S.~Santra and K.~Jacobs,
\newblock \emph{Ability of {{Markovian}} master equations to model quantum
  computers and other systems under broadband control},
\newblock Phys. Rev. Research \textbf{2}(1), 013049 (2020),
\newblock \doi{10.1103/PhysRevResearch.2.013049}.

\bibitem{leghtasDevoret2015}
Z.~Leghtas, S.~Touzard, I.~M. Pop, A.~Kou, B.~Vlastakis, A.~Petrenko, K.~M.
  Sliwa, A.~Narla, S.~Shankar, M.~J. Hatridge, M.~Reagor, L.~Frunzio
  \emph{et~al.},
\newblock \emph{Confining the state of light to a quantum manifold by
  engineered two-photon loss},
\newblock Science \textbf{347}(6224), 853 (2015),
\newblock \doi{10.1126/science.aaa2085}.

\bibitem{giovannettiMaccone2011}
V.~Giovannetti, S.~Lloyd and L.~Maccone,
\newblock \emph{Advances in quantum metrology},
\newblock Nature Photon \textbf{5}(4), 222 (2011),
\newblock \doi{10.1038/nphoton.2011.35}.

\bibitem{vinjanampathyAnders2016a}
S.~Vinjanampathy and J.~Anders,
\newblock \emph{Quantum thermodynamics},
\newblock Contemporary Physics \textbf{57}(4), 545 (2016),
\newblock \doi{10.1080/00107514.2016.1201896}.

\bibitem{komarLukin2014}
P.~K{\'o}m{\'a}r, E.~M. Kessler, M.~Bishof, L.~Jiang, A.~S. S{\o}rensen, J.~Ye
  and M.~D. Lukin,
\newblock \emph{A quantum network of clocks},
\newblock Nature Phys \textbf{10}(8), 582 (2014),
\newblock \doi{10.1038/nphys3000}.

\bibitem{gehringvanderZant2019}
P.~Gehring, J.~M. Thijssen and H.~S.~J. {van der Zant},
\newblock \emph{Single-molecule quantum-transport phenomena in break
  junctions},
\newblock Nat Rev Phys \textbf{1}(6), 381 (2019),
\newblock \doi{10.1038/s42254-019-0055-1}.

\bibitem{hanggiBorkovec1990}
P.~H{\"a}nggi, P.~Talkner and M.~Borkovec,
\newblock \emph{Reaction-rate theory: Fifty years after {{Kramers}}},
\newblock Rev. Mod. Phys. \textbf{62}(2), 251 (1990),
\newblock \doi{10.1103/RevModPhys.62.251}.

\bibitem{wolynesWolynes1981}
P.~G. Wolynes,
\newblock \emph{Quantum {{Theory}} of {{Activated Events}} in {{Condensed
  Phases}}},
\newblock Phys. Rev. Lett. \textbf{47}(13), 968 (1981),
\newblock \doi{10.1103/PhysRevLett.47.968}.

\bibitem{vothMiller1989}
G.~A. Voth, D.~Chandler and W.~H. Miller,
\newblock \emph{Rigorous formulation of quantum transition state theory and its
  dynamical corrections},
\newblock J. Chem. Phys. \textbf{91}(12), 7749 (1989),
\newblock \doi{10.1063/1.457242}.

\bibitem{ianconescuPollak2019}
R.~Ianconescu and E.~Pollak,
\newblock \emph{Activated quantum diffusion in a periodic potential above the
  crossover temperature},
\newblock J. Chem. Phys. \textbf{151}(2), 024703 (2019),
\newblock \doi{10.1063/1.5100010}.

\bibitem{engelFleming2007}
G.~S. Engel, T.~R. Calhoun, E.~L. Read, T.-K. Ahn, T.~Man{\v c}al, Y.-C. Cheng,
  R.~E. Blankenship and G.~R. Fleming,
\newblock \emph{Evidence for wavelike energy transfer through quantum coherence
  in photosynthetic systems},
\newblock Nature \textbf{446}(7137), 782 (2007),
\newblock \doi{10.1038/nature05678}.

\bibitem{panitchayangkoonEngel2010}
G.~Panitchayangkoon, D.~Hayes, K.~A. Fransted, J.~R. Caram, E.~Harel, J.~Wen,
  R.~E. Blankenship and G.~S. Engel,
\newblock \emph{Long-lived quantum coherence in photosynthetic complexes at
  physiological temperature},
\newblock PNAS \textbf{107}(29), 12766 (2010).

\bibitem{colliniScholes2010}
E.~Collini, C.~Y. Wong, K.~E. Wilk, P.~M.~G. Curmi, P.~Brumer and G.~D.
  Scholes,
\newblock \emph{Coherently wired light-harvesting in photosynthetic marine
  algae at ambient temperature},
\newblock Nature \textbf{463}(7281), 644 (2010),
\newblock \doi{10.1038/nature08811}.

\bibitem{blankenshipSayre2011}
R.~E. Blankenship, D.~M. Tiede, J.~Barber, G.~W. Brudvig, G.~Fleming,
  M.~Ghirardi, M.~R. Gunner, W.~Junge, D.~M. Kramer, A.~Melis, T.~A. Moore,
  C.~C. Moser \emph{et~al.},
\newblock \emph{Comparing {{Photosynthetic}} and {{Photovoltaic Efficiencies}}
  and {{Recognizing}} the {{Potential}} for {{Improvement}}},
\newblock Science \textbf{332}(6031), 805 (2011),
\newblock \doi{10.1126/science.1200165}.

\bibitem{lambertNori2013}
N.~Lambert, Y.-N. Chen, Y.-C. Cheng, C.-M. Li, G.-Y. Chen and F.~Nori,
\newblock \emph{Quantum biology},
\newblock Nature Phys \textbf{9}(1), 10 (2013),
\newblock \doi{10.1038/nphys2474}.

\bibitem{caoZigmantas2020}
J.~Cao, R.~J. Cogdell, D.~F. Coker, H.-G. Duan, J.~Hauer,
  U.~Kleinekath{\"o}fer, T.~L.~C. Jansen, T.~Man{\v c}al, R.~J.~D. Miller,
  J.~P. Ogilvie, V.~I. Prokhorenko, T.~Renger \emph{et~al.},
\newblock \emph{Quantum biology revisited},
\newblock Sci. Adv. \textbf{6}(14), eaaz4888 (2020),
\newblock \doi{10.1126/sciadv.aaz4888}.

\bibitem{kapitSimon2014a}
E.~Kapit, M.~Hafezi and S.~H. Simon,
\newblock \emph{Induced {{Self}}-{{Stabilization}} in {{Fractional Quantum Hall
  States}} of {{Light}}},
\newblock Phys. Rev. X \textbf{4}(3), 031039 (2014),
\newblock \doi{10.1103/PhysRevX.4.031039}.

\bibitem{maSchuster2019}
R.~Ma, B.~Saxberg, C.~Owens, N.~Leung, Y.~Lu, J.~Simon and D.~I. Schuster,
\newblock \emph{A dissipatively stabilized {{Mott}} insulator of photons},
\newblock Nature \textbf{566}(7742), 51 (2019),
\newblock \doi{10.1038/s41586-019-0897-9}.

\bibitem{puertasmartinezRoch2019a}
J.~Puertas~Mart{\'i}nez, S.~L{\'e}ger, N.~Gheeraert, R.~Dassonneville,
  L.~Planat, F.~Foroughi, Y.~Krupko, O.~Buisson, C.~Naud, W.~{Hasch-Guichard},
  S.~Florens, I.~Snyman \emph{et~al.},
\newblock \emph{A tunable {{Josephson}} platform to explore many-body quantum
  optics in circuit-{{QED}}},
\newblock npj Quantum Inf \textbf{5}(1), 19 (2019),
\newblock \doi{10.1038/s41534-018-0104-0}.

\bibitem{carusottoSimon2020a}
I.~Carusotto, A.~A. Houck, A.~J. Koll{\'a}r, P.~Roushan, D.~I. Schuster and
  J.~Simon,
\newblock \emph{Photonic materials in circuit quantum electrodynamics},
\newblock Nat. Phys. \textbf{16}(3), 268 (2020),
\newblock \doi{10.1038/s41567-020-0815-y}.

\bibitem{scullyZubairy1997}
M.~O. Scully and M.~S. Zubairy,
\newblock \emph{Quantum {{Optics}}},
\newblock {Cambridge University Press}, {Cambridge},
\newblock ISBN 978-0-511-81399-3 (1997).

\bibitem{daviesDavies1974}
E.~B. Davies,
\newblock \emph{Markovian master equations},
\newblock Commun.Math. Phys. \textbf{39}(2), 91 (1974),
\newblock \doi{10.1007/BF01608389}.

\bibitem{lindbladLindblad1976a}
G.~Lindblad,
\newblock \emph{On the generators of quantum dynamical semigroups},
\newblock Communications in Mathematical Physics \textbf{48}(2), 119 (1976),
\newblock \doi{10.1007/BF01608499}.

\bibitem{dumckeSpohn1979}
R.~D{\"u}mcke and H.~Spohn,
\newblock \emph{The proper form of the generator in the weak coupling limit},
\newblock Z Physik B \textbf{34}(4), 419 (1979),
\newblock \doi{10.1007/BF01325208}.

\bibitem{nathanRudner2020}
F.~Nathan and M.~S. Rudner,
\newblock \emph{Universal {{Lindblad}} equation for open quantum systems},
\newblock Phys. Rev. B \textbf{102}(11), 115109 (2020),
\newblock \doi{10.1103/PhysRevB.102.115109}.

\bibitem{mozgunovLidar2020}
E.~Mozgunov and D.~Lidar,
\newblock \emph{Completely positive master equation for arbitrary driving and
  small level spacing},
\newblock Quantum \textbf{4}, 227 (2020),
\newblock \doi{10.22331/q-2020-02-06-227}.

\bibitem{mccauleyJacobs2020a}
G.~McCauley, B.~Cruikshank, D.~I. Bondar and K.~Jacobs,
\newblock \emph{Accurate {{Lindblad}}-form master equation for weakly damped
  quantum systems across all regimes},
\newblock npj Quantum Inf \textbf{6}(1), 1 (2020),
\newblock \doi{10.1038/s41534-020-00299-6}.

\bibitem{davidovicDavidovic2020}
D.~Davidovi{\'c},
\newblock \emph{Completely {{Positive}}, {{Simple}}, and {{Possibly Highly
  Accurate Approximation}} of the {{Redfield Equation}}},
\newblock Quantum \textbf{4}, 326 (2020),
\newblock \doi{10.22331/q-2020-09-21-326}.

\bibitem{trushechkinTrushechkin2021}
A.~Trushechkin,
\newblock \emph{Unified {{GKLS}} quantum master equation of weak-coupling limit
  type},
\newblock arXiv:2103.12042 [math-ph, physics:quant-ph]  (2021).

\bibitem{becker2021lindbladian}
T.~Becker, L.-N. Wu and A.~Eckardt,
\newblock \emph{Lindbladian approximation beyond ultra-weak coupling} (2021),
  \eprint{2012.14208}.

\bibitem{dalibardMolmer1992}
J.~Dalibard, Y.~Castin and K.~M{\o}lmer,
\newblock \emph{Wave-function approach to dissipative processes in quantum
  optics},
\newblock Phys. Rev. Lett. \textbf{68}(5), 580 (1992),
\newblock \doi{10.1103/PhysRevLett.68.580}.

\bibitem{dumRitsch1992}
R.~Dum, P.~Zoller and H.~Ritsch,
\newblock \emph{Monte {{Carlo}} simulation of the atomic master equation for
  spontaneous emission},
\newblock Phys. Rev. A \textbf{45}(7), 4879 (1992),
\newblock \doi{10.1103/PhysRevA.45.4879}.

\bibitem{gardinerZoller2004}
C.~W. Gardiner and P.~Zoller,
\newblock \emph{Quantum Noise: A Handbook of {{Markovian}} and
  Non-{{Markovian}} Quantum Stochastic Methods with Applications to Quantum
  Optics},
\newblock Springer Series in Synergetics. {Springer}, {Berlin ; New York}, 3rd
  ed edn.,
\newblock ISBN 978-3-540-22301-6 (2004).

\bibitem{espositoGalperin2009}
M.~Esposito and M.~Galperin,
\newblock \emph{Transport in molecular states language: {{Generalized}} quantum
  master equation approach},
\newblock Phys. Rev. B \textbf{79}(20), 205303 (2009),
\newblock \doi{10.1103/PhysRevB.79.205303}.

\bibitem{espositoGalperin2010}
M.~Esposito and M.~Galperin,
\newblock \emph{Self-{{Consistent Quantum Master Equation Approach}} to
  {{Molecular Transport}}},
\newblock J. Phys. Chem. C \textbf{114}(48), 20362 (2010),
\newblock \doi{10.1021/jp103369s}.

\bibitem{jinYan2014}
J.~Jin, J.~Li, Y.~Liu, X.-Q. Li and Y.~Yan,
\newblock \emph{Improved master equation approach to quantum transport: {{From
  Born}} to self-consistent {{Born}} approximation},
\newblock The Journal of Chemical Physics \textbf{140}(24), 244111 (2014),
\newblock \doi{10.1063/1.4884390}.

\bibitem{sowaGauger2018}
J.~K. Sowa, J.~A. Mol, G.~A.~D. Briggs and E.~M. Gauger,
\newblock \emph{Beyond {{Marcus}} theory and the {{Landauer}}-{{B\"uttiker}}
  approach in molecular junctions: {{A}} unified framework},
\newblock The Journal of Chemical Physics \textbf{149}(15), 154112 (2018),
\newblock \doi{10.1063/1.5049537}.

\bibitem{sowaGauger2020}
J.~K. Sowa, N.~Lambert, T.~Seideman and E.~M. Gauger,
\newblock \emph{Beyond {{Marcus}} theory and the
  {{Landauer}}\textendash{{B\"uttiker}} approach in molecular junctions.
  {{II}}. {{A}} self-consistent {{Born}} approach},
\newblock J. Chem. Phys. \textbf{152}(6), 064103 (2020),
\newblock \doi{10.1063/1.5143146}.

\bibitem{groblacherEisert2015}
S.~Gr{\"o}blacher, A.~Trubarov, N.~Prigge, G.~D. Cole, M.~Aspelmeyer and
  J.~Eisert,
\newblock \emph{Observation of non-{{Markovian}} micromechanical {{Brownian}}
  motion},
\newblock Nat Commun \textbf{6}(1), 7606 (2015),
\newblock \doi{10.1038/ncomms8606}.

\bibitem{madsenLodahl2011}
K.~H. Madsen, S.~Ates, T.~{Lund-Hansen}, A.~L{\"o}ffler, S.~Reitzenstein,
  A.~Forchel and P.~Lodahl,
\newblock \emph{Observation of {{Non-Markovian Dynamics}} of a {{Single Quantum
  Dot}} in a {{Micropillar Cavity}}},
\newblock Physical Review Letters \textbf{106}(23), 233601 (2011),
\newblock \doi{10.1103/PhysRevLett.106.233601}.

\bibitem{miPetta2017}
X.~Mi, J.~V. Cady, D.~M. Zajac, P.~W. Deelman and J.~R. Petta,
\newblock \emph{Strong coupling of a single electron in silicon to a microwave
  photon},
\newblock Science \textbf{355}(6321), 156 (2017),
\newblock \doi{10.1126/science.aal2469}.

\bibitem{whiteModi2020}
G.~a.~L. White, C.~D. Hill, F.~A. Pollock, L.~C.~L. Hollenberg and K.~Modi,
\newblock \emph{Demonstration of non-{{Markovian}} process characterisation and
  control on a quantum processor},
\newblock Nature Communications \textbf{11}(1), 6301 (2020),
\newblock \doi{10.1038/s41467-020-20113-3}.

\bibitem{papicdeVega2023}
M.~Papi{\v c}, A.~Auer and I.~{de Vega},
\newblock \emph{Fast {{Estimation}} of {{Physical Error Contributions}} of
  {{Quantum Gates}}} (2023).

\bibitem{paladinoAltshuler2014}
E.~Paladino, Y.~M. Galperin, G.~Falci and B.~L. Altshuler,
\newblock \emph{\$\textbackslash mathbsf\{1\}/\textbackslash mathbsfit\{f\}\$
  noise: {{Implications}} for solid-state quantum information},
\newblock Reviews of Modern Physics \textbf{86}(2), 361 (2014),
\newblock \doi{10.1103/RevModPhys.86.361}.

\bibitem{burnettBylander2019}
J.~J. Burnett, A.~Bengtsson, M.~Scigliuzzo, D.~Niepce, M.~Kudra, P.~Delsing and
  J.~Bylander,
\newblock \emph{Decoherence benchmarking of superconducting qubits},
\newblock npj Quantum Information \textbf{5}(1), 1 (2019),
\newblock \doi{10.1038/s41534-019-0168-5}.

\bibitem{rowerOliver2023}
D.~A. Rower, L.~Ateshian, L.~H. Li, M.~Hays, D.~Bluvstein, L.~Ding, B.~Kannan,
  A.~Almanakly, J.~Braum{\"u}ller, D.~K. Kim, A.~Melville, B.~M. Niedzielski
  \emph{et~al.},
\newblock \emph{Evolution of 1 / f {{Flux Noise}} in {{Superconducting Qubits}}
  with {{Weak Magnetic Fields}}},
\newblock Physical Review Letters \textbf{130}(22), 220602 (2023),
\newblock \doi{10.1103/PhysRevLett.130.220602}.

\bibitem{hansonAwschalom2008}
R.~Hanson, V.~V. Dobrovitski, A.~E. Feiguin, O.~Gywat and D.~D. Awschalom,
\newblock \emph{Coherent {{Dynamics}} of a {{Single Spin Interacting}} with an
  {{Adjustable Spin Bath}}},
\newblock Science \textbf{320}(5874), 352 (2008),
\newblock \doi{10.1126/science.1155400}.

\bibitem{lebratEsslinger2018}
M.~Lebrat, P.~Gri{\v s}ins, D.~Husmann, S.~H{\"a}usler, L.~Corman,
  T.~Giamarchi, J.-P. Brantut and T.~Esslinger,
\newblock \emph{Band and {{Correlated Insulators}} of {{Cold Fermions}} in a
  {{Mesoscopic Lattice}}},
\newblock Phys. Rev. X \textbf{8}(1), 011053 (2018),
\newblock \doi{10.1103/PhysRevX.8.011053}.

\bibitem{liuPiilo2011}
B.-H. Liu, L.~Li, Y.-F. Huang, C.-F. Li, G.-C. Guo, E.-M. Laine, H.-P. Breuer
  and J.~Piilo,
\newblock \emph{Experimental control of the transition from {{Markovian}} to
  non-{{Markovian}} dynamics of open quantum systems},
\newblock Nature Phys \textbf{7}(12), 931 (2011),
\newblock \doi{10.1038/nphys2085}.

\bibitem{maierRoos2019}
C.~Maier, T.~Brydges, P.~Jurcevic, N.~Trautmann, C.~Hempel, B.~P. Lanyon,
  P.~Hauke, R.~Blatt and C.~F. Roos,
\newblock \emph{Environment-{{Assisted Quantum Transport}} in a 10-qubit
  {{Network}}},
\newblock Phys. Rev. Lett. \textbf{122}(5), 050501 (2019),
\newblock \doi{10.1103/PhysRevLett.122.050501}.

\bibitem{recatiZoller2005}
A.~Recati, P.~O. Fedichev, W.~Zwerger, J.~{von Delft} and P.~Zoller,
\newblock \emph{Atomic {{Quantum Dots Coupled}} to a {{Reservoir}} of a
  {{Superfluid Bose}}-{{Einstein Condensate}}},
\newblock Phys. Rev. Lett. \textbf{94}(4), 040404 (2005),
\newblock \doi{10.1103/PhysRevLett.94.040404}.

\bibitem{puertasmartinezRoch2019}
J.~Puertas~Mart{\'i}nez, S.~L{\'e}ger, N.~Gheeraert, R.~Dassonneville,
  L.~Planat, F.~Foroughi, Y.~Krupko, O.~Buisson, C.~Naud, W.~{Hasch-Guichard},
  S.~Florens, I.~Snyman \emph{et~al.},
\newblock \emph{A tunable {{Josephson}} platform to explore many-body quantum
  optics in circuit-{{QED}}},
\newblock npj Quantum Inf \textbf{5}(1), 19 (2019),
\newblock \doi{10.1038/s41534-018-0104-0}.

\bibitem{devega2017dynamics}
I.~de~Vega and D.~Alonso,
\newblock \emph{Dynamics of non-markovian open quantum systems},
\newblock Rev. Mod. Phys. \textbf{89}, 015001 (2017),
\newblock \doi{10.1103/RevModPhys.89.015001}.

\bibitem{liPiilo2020}
C.-F. Li, G.-C. Guo and J.~Piilo,
\newblock \emph{Non-{{Markovian}} quantum dynamics: {{What}} is it good for?},
\newblock EPL \textbf{128}(3), 30001 (2020),
\newblock \doi{10.1209/0295-5075/128/30001}.

\bibitem{chinPlenio2012}
A.~W. Chin, S.~F. Huelga and M.~B. Plenio,
\newblock \emph{Quantum {{Metrology}} in {{Non}}-{{Markovian Environments}}},
\newblock Phys. Rev. Lett. \textbf{109}(23), 233601 (2012),
\newblock \doi{10.1103/PhysRevLett.109.233601}.

\bibitem{alickiHorodecki2002}
R.~Alicki, M.~Horodecki, P.~Horodecki and R.~Horodecki,
\newblock \emph{Dynamical description of quantum computing: {{Generic}}
  nonlocality of quantum noise},
\newblock Phys. Rev. A \textbf{65}(6), 062101 (2002),
\newblock \doi{10.1103/PhysRevA.65.062101}.

\bibitem{dongSun2018}
Y.~Dong, Y.~Zheng, S.~Li, C.-C. Li, X.-D. Chen, G.-C. Guo and F.-W. Sun,
\newblock \emph{Non-{{Markovianity}}-assisted high-fidelity
  {{Deutsch}}\textendash{{Jozsa}} algorithm in diamond},
\newblock npj Quantum Inf \textbf{4}(1), 1 (2018),
\newblock \doi{10.1038/s41534-017-0053-z}.

\bibitem{shabaniLidar2005}
A.~Shabani and D.~A. Lidar,
\newblock \emph{Completely positive post-{{Markovian}} master equation via a
  measurement approach},
\newblock Phys. Rev. A \textbf{71}(2), 020101 (2005),
\newblock \doi{10.1103/PhysRevA.71.020101}.

\bibitem{maniscalcoPetruccione2006}
S.~Maniscalco and F.~Petruccione,
\newblock \emph{Non-{{Markovian}} dynamics of a qubit},
\newblock Phys. Rev. A \textbf{73}(1), 012111 (2006),
\newblock \doi{10.1103/PhysRevA.73.012111}.

\bibitem{chruscinskiKossakowski2010a}
D.~Chru{\'s}ci{\'n}ski and A.~Kossakowski,
\newblock \emph{Non-{{Markovian Quantum Dynamics}}: {{Local}} versus
  {{Nonlocal}}},
\newblock Phys. Rev. Lett. \textbf{104}(7), 070406 (2010),
\newblock \doi{10.1103/PhysRevLett.104.070406}.

\bibitem{schoellerSchon1994}
H.~Schoeller and G.~Sch{\"o}n,
\newblock \emph{Mesoscopic quantum transport: {{Resonant}} tunneling in the
  presence of a strong {{Coulomb}} interaction},
\newblock Phys. Rev. B \textbf{50}(24), 18436 (1994),
\newblock \doi{10.1103/PhysRevB.50.18436}.

\bibitem{schoellerSchoeller2009}
H.~Schoeller,
\newblock \emph{A perturbative nonequilibrium renormalization group method for
  dissipative quantum mechanics: {{Real}}-time {{RG}} in frequency space
  ({{RTRG}}-{{FS}})},
\newblock Eur. Phys. J. Spec. Top. \textbf{168}(1), 179 (2009),
\newblock \doi{10.1140/epjst/e2009-00962-3}.

\bibitem{karlewskiMarthaler2014}
C.~Karlewski and M.~Marthaler,
\newblock \emph{Time-local master equation connecting the {{Born}} and
  {{Markov}} approximations},
\newblock Phys. Rev. B \textbf{90}(10), 104302 (2014),
\newblock \doi{10.1103/PhysRevB.90.104302}.

\bibitem{mullerStace2017}
C.~M{\"u}ller and T.~M. Stace,
\newblock \emph{Deriving {{Lindblad}} master equations with {{Keldysh}}
  diagrams: {{Correlated}} gain and loss in higher order perturbation theory},
\newblock Phys. Rev. A \textbf{95}(1), 013847 (2017),
\newblock \doi{10.1103/PhysRevA.95.013847}.

\bibitem{lindnerSchoeller2018}
C.~J. Lindner and H.~Schoeller,
\newblock \emph{Dissipative quantum mechanics beyond the {{Bloch}}-{{Redfield}}
  approximation: {{A}} consistent weak-coupling expansion of the {{Ohmic}} spin
  boson model at arbitrary bias},
\newblock Phys. Rev. B \textbf{98}(11), 115425 (2018),
\newblock \doi{10.1103/PhysRevB.98.115425}.

\bibitem{kleinherbersSchutzhold2020}
E.~Kleinherbers, N.~Szpak, J.~K{\"o}nig and R.~Sch{\"u}tzhold,
\newblock \emph{Relaxation dynamics in a {{Hubbard}} dimer coupled to fermionic
  baths: {{Phenomenological}} description and its microscopic foundation},
\newblock Phys. Rev. B \textbf{101}(12), 125131 (2020),
\newblock \doi{10.1103/PhysRevB.101.125131}.

\bibitem{mulhbacher2008realtime}
L.~M\"uhlbacher and E.~Rabani,
\newblock \emph{Real-time path integral approach to nonequilibrium many-body
  quantum systems},
\newblock Phys. Rev. Lett. \textbf{100}, 176403 (2008),
\newblock \doi{10.1103/PhysRevLett.100.176403}.

\bibitem{schiro2009realtime}
M.~Schir\'o and M.~Fabrizio,
\newblock \emph{Real-time diagrammatic monte carlo for nonequilibrium quantum
  transport},
\newblock Phys. Rev. B \textbf{79}, 153302 (2009),
\newblock \doi{10.1103/PhysRevB.79.153302}.

\bibitem{werner2009diagrammatic}
P.~Werner, T.~Oka and A.~J. Millis,
\newblock \emph{Diagrammatic monte carlo simulation of nonequilibrium systems},
\newblock Phys. Rev. B \textbf{79}, 035320 (2009),
\newblock \doi{10.1103/PhysRevB.79.035320}.

\bibitem{cohen2014green}
G.~Cohen, E.~Gull, D.~R. Reichman and A.~J. Millis,
\newblock \emph{Green's functions from real-time bold-line monte carlo
  calculations: Spectral properties of the nonequilibrium anderson impurity
  model},
\newblock Phys. Rev. Lett. \textbf{112}, 146802 (2014),
\newblock \doi{10.1103/PhysRevLett.112.146802}.

\bibitem{cohen2015taming}
G.~Cohen, E.~Gull, D.~R. Reichman and A.~J. Millis,
\newblock \emph{Taming the dynamical sign problem in real-time evolution of
  quantum many-body problems},
\newblock Phys. Rev. Lett. \textbf{115}, 266802 (2015),
\newblock \doi{10.1103/PhysRevLett.115.266802}.

\bibitem{chenReichman2017b}
H.-T. Chen, G.~Cohen and D.~R. Reichman,
\newblock \emph{Inchworm {{Monte Carlo}} for exact non-adiabatic dynamics.
  {{I}}. {{Theory}} and algorithms},
\newblock J. Chem. Phys. \textbf{146}(5), 054105 (2017),
\newblock \doi{10.1063/1.4974328}.

\bibitem{chenReichman2017g}
H.-T. Chen, G.~Cohen and D.~R. Reichman,
\newblock \emph{Inchworm {{Monte Carlo}} for exact non-adiabatic dynamics.
  {{II}}. {{Benchmarks}} and comparison with established methods},
\newblock J. Chem. Phys. \textbf{146}(5), 054106 (2017),
\newblock \doi{10.1063/1.4974329}.

\bibitem{jinYan2008}
J.~Jin, X.~Zheng and Y.~Yan,
\newblock \emph{Exact dynamics of dissipative electronic systems and quantum
  transport: {{Hierarchical}} equations of motion approach},
\newblock The Journal of Chemical Physics \textbf{128}(23), 234703 (2008),
\newblock \doi{10.1063/1.2938087}.

\bibitem{hartleMillis2013}
R.~H{\"a}rtle, G.~Cohen, D.~R. Reichman and A.~J. Millis,
\newblock \emph{Decoherence and lead-induced interdot coupling in
  nonequilibrium electron transport through interacting quantum dots: {{A}}
  hierarchical quantum master equation approach},
\newblock Phys. Rev. B \textbf{88}(23), 235426 (2013),
\newblock \doi{10.1103/PhysRevB.88.235426}.

\bibitem{tanimuraTanimura2020}
Y.~Tanimura,
\newblock \emph{Numerically ``exact'' approach to open quantum dynamics:
  {{The}} hierarchical equations of motion ({{HEOM}})},
\newblock J. Chem. Phys. \textbf{153}(2), 020901 (2020),
\newblock \doi{10.1063/5.0011599}.

\bibitem{diosiStrunz1997}
L.~Di{\'o}si and W.~T. Strunz,
\newblock \emph{The non-{{Markovian}} stochastic {{Schr\"odinger}} equation for
  open systems},
\newblock Physics Letters A \textbf{235}(6), 569 (1997),
\newblock \doi{10.1016/S0375-9601(97)00717-2}.

\bibitem{diosiStrunz1998}
L.~Di{\'o}si, N.~Gisin and W.~T. Strunz,
\newblock \emph{Non-{{Markovian}} quantum state diffusion},
\newblock Phys. Rev. A \textbf{58}(3), 1699 (1998),
\newblock \doi{10.1103/PhysRevA.58.1699}.

\bibitem{gaspardNagaoka1999}
P.~Gaspard and M.~Nagaoka,
\newblock \emph{Non-{{Markovian}} stochastic {{Schr\"odinger}} equation},
\newblock J. Chem. Phys. \textbf{111}(13), 5676 (1999),
\newblock \doi{10.1063/1.479868}.

\bibitem{jingYu2012}
J.~Jing, X.~Zhao, J.~Q. You and T.~Yu,
\newblock \emph{Time-local quantum-state-diffusion equation for multilevel
  quantum systems},
\newblock Phys. Rev. A \textbf{85}(4), 042106 (2012),
\newblock \doi{10.1103/PhysRevA.85.042106}.

\bibitem{suessStrunz2014}
D.~Suess, A.~Eisfeld and W.~T. Strunz,
\newblock \emph{Hierarchy of {{Stochastic Pure States}} for {{Open Quantum
  System Dynamics}}},
\newblock Phys. Rev. Lett. \textbf{113}(15), 150403 (2014),
\newblock \doi{10.1103/PhysRevLett.113.150403}.

\bibitem{thoennissAbanin2022}
J.~Thoenniss, M.~Sonner, A.~Lerose and D.~A. Abanin,
\newblock \emph{An efficient method for quantum impurity problems out of
  equilibrium} (2022).

\bibitem{strathearnLovett2018}
A.~Strathearn, P.~Kirton, D.~Kilda, J.~Keeling and B.~W. Lovett,
\newblock \emph{Efficient non-{{Markovian}} quantum dynamics using
  time-evolving matrix product operators},
\newblock Nat Commun \textbf{9}(1), 3322 (2018),
\newblock \doi{10.1038/s41467-018-05617-3}.

\bibitem{altland_simons_2010}
A.~Altland and B.~D. Simons,
\newblock \emph{Condensed Matter Field Theory},
\newblock Cambridge University Press, 2 edn.,
\newblock \doi{10.1017/CBO9780511789984} (2010).

\bibitem{bruusFlensberg2004}
H.~Bruus and K.~Flensberg,
\newblock \emph{Many-Body Quantum Theory in Condensed Matter Physics : An
  Introduction},
\newblock {Oxford University Press},
\newblock ISBN 978-0-19-856633-5 (2004).

\bibitem{bickersBickers1987b}
N.~E. Bickers,
\newblock \emph{Review of techniques in the large- {{N}} expansion for dilute
  magnetic alloys},
\newblock Rev. Mod. Phys. \textbf{59}(4), 845 (1987),
\newblock \doi{10.1103/RevModPhys.59.845}.

\bibitem{muller-hartmannMuller-Hartmann1984}
E.~{M{\"u}ller-Hartmann},
\newblock \emph{Self-consistent perturbation theory of the anderson model:
  {{Ground}} state properties},
\newblock Z. Physik B - Condensed Matter \textbf{57}(4), 281 (1984),
\newblock \doi{10.1007/BF01470417}.

\bibitem{nordlander1999how}
P.~Nordlander, M.~Pustilnik, Y.~Meir, N.~S. Wingreen and D.~C. Langreth,
\newblock \emph{How long does it take for the kondo effect to develop?},
\newblock Phys. Rev. Lett. \textbf{83}, 808 (1999),
\newblock \doi{10.1103/PhysRevLett.83.808}.

\bibitem{ecksteinWerner2010a}
M.~Eckstein and P.~Werner,
\newblock \emph{Nonequilibrium dynamical mean-field calculations based on the
  noncrossing approximation and its generalizations},
\newblock Phys. Rev. B \textbf{82}(11), 115115 (2010),
\newblock \doi{10.1103/PhysRevB.82.115115}.

\bibitem{chen2016anderson}
H.-T. Chen, G.~Cohen, A.~J. Millis and D.~R. Reichman,
\newblock \emph{Anderson-holstein model in two flavors of the noncrossing
  approximation},
\newblock Phys. Rev. B \textbf{93}, 174309 (2016),
\newblock \doi{10.1103/PhysRevB.93.174309}.

\bibitem{meir1993low}
Y.~Meir, N.~S. Wingreen and P.~A. Lee,
\newblock \emph{Low-temperature transport through a quantum dot: The anderson
  model out of equilibrium},
\newblock Phys. Rev. Lett. \textbf{70}, 2601 (1993),
\newblock \doi{10.1103/PhysRevLett.70.2601}.

\bibitem{pruschkeJarrell1993}
T.~Pruschke, D.~L. Cox and M.~Jarrell,
\newblock \emph{Hubbard model at infinite dimensions: {{Thermodynamic}} and
  transport properties},
\newblock Phys. Rev. B \textbf{47}(7), 3553 (1993),
\newblock \doi{10.1103/PhysRevB.47.3553}.

\bibitem{erpenbeck2021revealing}
A.~Erpenbeck, E.~Gull and G.~Cohen,
\newblock \emph{Revealing strong correlations in higher-order transport
  statistics: A noncrossing approximation approach},
\newblock Phys. Rev. B \textbf{103}, 125431 (2021),
\newblock \doi{10.1103/PhysRevB.103.125431}.

\bibitem{hartle2013decoherence}
R.~H\"artle, G.~Cohen, D.~R. Reichman and A.~J. Millis,
\newblock \emph{Decoherence and lead-induced interdot coupling in
  nonequilibrium electron transport through interacting quantum dots: A
  hierarchical quantum master equation approach},
\newblock Phys. Rev. B \textbf{88}, 235426 (2013),
\newblock \doi{10.1103/PhysRevB.88.235426}.

\bibitem{schiroScarlatella2019}
M.~Schir{\'o} and O.~Scarlatella,
\newblock \emph{Quantum impurity models coupled to {{Markovian}} and
  non-{{Markovian}} baths},
\newblock The Journal of Chemical Physics \textbf{151}(4), 044102 (2019),
\newblock \doi{10.1063/1.5100157}.

\bibitem{breuerPetruccione2007}
H.~P. Breuer and F.~Petruccione,
\newblock \emph{The {{Theory}} of {{Open Quantum Systems}}}, vol. 9780199213,
\newblock {OUP Oxford}, 1st ed edn.,
\newblock ISBN 978-0-19-170634-9,
\newblock \doi{10.1093/acprof:oso/9780199213900.001.0001} (2007).

\bibitem{hartmannStrunz2020}
R.~Hartmann and W.~T. Strunz,
\newblock \emph{Accuracy assessment of perturbative master equations:
  {{Embracing}} nonpositivity},
\newblock Physical Review A \textbf{101}(1), 012103 (2020),
\newblock \doi{10.1103/PhysRevA.101.012103}.

\bibitem{scarlatellaSchiro2021b}
O.~Scarlatella, A.~A. Clerk, R.~Fazio and M.~Schir{\'o},
\newblock \emph{Dynamical {{Mean}}-{{Field Theory}} for {{Markovian Open
  Quantum Many}}-{{Body Systems}}},
\newblock Phys. Rev. X \textbf{11}(3), 031018 (2021),
\newblock \doi{10.1103/PhysRevX.11.031018}.

\bibitem{stefanucci2013nonequilibrium}
G.~Stefanucci and R.~{van Leeuwen},
\newblock \emph{Nonequilibrium Many-Body Theory of Quantum Systems: {{A}}
  Modern Introduction},
\newblock {Cambridge University Press},
\newblock ISBN 978-0-521-76617-3 (2013).

\bibitem{gullMillis2010}
E.~Gull, D.~R. Reichman and A.~J. Millis,
\newblock \emph{Bold-line diagrammatic {{Monte Carlo}} method: {{General}}
  formulation and application to expansion around the noncrossing
  approximation},
\newblock Phys. Rev. B \textbf{82}(7), 075109 (2010),
\newblock \doi{10.1103/PhysRevB.82.075109}.

\bibitem{kimEckstein2023}
A.~J. Kim, K.~Lenk, J.~Li, P.~Werner and M.~Eckstein,
\newblock \emph{Vertex-{{Based Diagrammatic Treatment}} of
  {{Light-Matter-Coupled Systems}}},
\newblock Physical Review Letters \textbf{130}(3), 036901 (2023),
\newblock \doi{10.1103/PhysRevLett.130.036901}.

\bibitem{lidarWhaley2001}
D.~A. Lidar, Z.~Bihary and K.~B. Whaley,
\newblock \emph{From completely positive maps to the quantum {{Markovian}}
  semigroup master equation},
\newblock Chemical Physics \textbf{268}(1), 35 (2001),
\newblock \doi{10.1016/S0301-0104(01)00330-5}.

\bibitem{majenzLidar2013}
C.~Majenz, T.~Albash, H.-P. Breuer and D.~A. Lidar,
\newblock \emph{Coarse graining can beat the rotating-wave approximation in
  quantum {{Markovian}} master equations},
\newblock PHYSICAL REVIEW A p.~16 (2013).

\bibitem{whitneyWhitney2008}
R.~S. Whitney,
\newblock \emph{Staying positive: Going beyond {{Lindblad}} with perturbative
  master equations},
\newblock J. Phys. A: Math. Theor. \textbf{41}(17), 175304 (2008),
\newblock \doi{10.1088/1751-8113/41/17/175304}.

\bibitem{aokiWerner2014}
H.~Aoki, N.~Tsuji, M.~Eckstein, M.~Kollar, T.~Oka and P.~Werner,
\newblock \emph{Nonequilibrium dynamical mean-field theory and its
  applications},
\newblock Reviews of Modern Physics \textbf{86}(2), 779 (2014),
\newblock \doi{10.1103/RevModPhys.86.779}.

\bibitem{georgesRozenberg1996}
A.~Georges, G.~Kotliar, W.~Krauth and M.~J. Rozenberg,
\newblock \emph{Dynamical mean-field theory of strongly correlated fermion
  systems and the limit of infinite dimensions},
\newblock Rev. Mod. Phys. \textbf{68}(1), 13 (1996),
\newblock \doi{10.1103/RevModPhys.68.13}.

\bibitem{verstraeteCirac2004}
F.~Verstraete, J.~J. {Garc{\'i}a-Ripoll} and J.~I. Cirac,
\newblock \emph{Matrix {{Product Density Operators}}: {{Simulation}} of
  {{Finite-Temperature}} and {{Dissipative Systems}}},
\newblock Physical Review Letters \textbf{93}(20), 207204 (2004),
\newblock \doi{10.1103/PhysRevLett.93.207204}.

\bibitem{zwolakVidal2004}
M.~Zwolak and G.~Vidal,
\newblock \emph{Mixed-{{State Dynamics}} in {{One-Dimensional Quantum Lattice
  Systems}}: {{A Time-Dependent Superoperator Renormalization Algorithm}}},
\newblock Physical Review Letters \textbf{93}(20), 207205 (2004),
\newblock \doi{10.1103/PhysRevLett.93.207205}.

\bibitem{flanniganDaley2021}
S.~Flannigan, F.~Damanet and A.~J. Daley,
\newblock \emph{Many-body quantum state diffusion for non-{{Markovian}}
  dynamics in strongly interacting systems},
\newblock ArXiv210806224 Quant-Ph  (2021).

\bibitem{blumeLuther1970}
M.~Blume, V.~J. Emery and A.~Luther,
\newblock \emph{Spin-{{Boson Systems}}: {{One}}-{{Dimensional Equivalents}} and
  the {{Kondo Problem}}},
\newblock Phys. Rev. Lett. \textbf{25}(7), 450 (1970),
\newblock \doi{10.1103/PhysRevLett.25.450}.

\bibitem{brayMoore1982}
A.~J. Bray and M.~A. Moore,
\newblock \emph{Influence of {{Dissipation}} on {{Quantum Coherence}}},
\newblock Phys. Rev. Lett. \textbf{49}(21), 1545 (1982),
\newblock \doi{10.1103/PhysRevLett.49.1545}.

\bibitem{chakravartyChakravarty1982}
S.~Chakravarty,
\newblock \emph{Quantum {{Fluctuations}} in the {{Tunneling}} between
  {{Superconductors}}},
\newblock Phys. Rev. Lett. \textbf{49}(9), 681 (1982),
\newblock \doi{10.1103/PhysRevLett.49.681}.

\bibitem{chakravartyLeggett1984}
S.~Chakravarty and A.~J. Leggett,
\newblock \emph{Dynamics of the {{Two}}-{{State System}} with {{Ohmic
  Dissipation}}},
\newblock Phys. Rev. Lett. \textbf{52}(1), 5 (1984),
\newblock \doi{10.1103/PhysRevLett.52.5}.

\bibitem{leggettZwerger1987}
A.~J. Leggett, S.~Chakravarty, A.~T. Dorsey, M.~P.~A. Fisher, A.~Garg and
  W.~Zwerger,
\newblock \emph{Dynamics of the dissipative two-state system},
\newblock Rev. Mod. Phys. \textbf{59}(1), 1 (1987),
\newblock \doi{10.1103/RevModPhys.59.1}.

\bibitem{keilSchoeller2001}
M.~Keil and H.~Schoeller,
\newblock \emph{Real-time renormalization-group analysis of the dynamics of the
  spin-boson model},
\newblock Phys. Rev. B \textbf{63}(18), 180302 (2001),
\newblock \doi{10.1103/PhysRevB.63.180302}.

\bibitem{andersSchiller2006}
F.~B. Anders and A.~Schiller,
\newblock \emph{Spin precession and real-time dynamics in the {{Kondo}} model:
  {{Time}}-dependent numerical renormalization-group study},
\newblock Phys. Rev. B \textbf{74}(24), 245113 (2006),
\newblock \doi{10.1103/PhysRevB.74.245113}.

\bibitem{andersVojta2007}
F.~B. Anders, R.~Bulla and M.~Vojta,
\newblock \emph{Equilibrium and {{Nonequilibrium Dynamics}} of the
  {{Sub}}-{{Ohmic Spin}}-{{Boson Model}}},
\newblock Phys. Rev. Lett. \textbf{98}(21), 210402 (2007),
\newblock \doi{10.1103/PhysRevLett.98.210402}.

\bibitem{bullaVojta2005}
R.~Bulla, H.-J. Lee, N.-H. Tong and M.~Vojta,
\newblock \emph{Numerical renormalization group for quantum impurities in a
  bosonic bath},
\newblock Phys. Rev. B \textbf{71}(4), 045122 (2005),
\newblock \doi{10.1103/PhysRevB.71.045122}.

\bibitem{vojtaVojta2012}
M.~Vojta,
\newblock \emph{Numerical renormalization group for the sub-{{Ohmic}}
  spin-boson model: {{A}} conspiracy of errors},
\newblock Phys. Rev. B \textbf{85}(11), 115113 (2012),
\newblock \doi{10.1103/PhysRevB.85.115113}.

\bibitem{amicoRibeiro2007}
L.~Amico, H.~Frahm, A.~Osterloh and G.~Ribeiro,
\newblock \emph{Integrable spin\textendash boson models descending from
  rational six-vertex models},
\newblock Nuclear Physics B \textbf{787}(3), 283 (2007),
\newblock \doi{10.1016/j.nuclphysb.2007.07.022}.

\bibitem{koppHur2007}
A.~Kopp and K.~L. Hur,
\newblock \emph{Universal and {{Measurable Entanglement Entropy}} in the
  {{Spin}}-{{Boson Model}}},
\newblock Phys. Rev. Lett. \textbf{98}(22), 220401 (2007),
\newblock \doi{10.1103/PhysRevLett.98.220401}.

\bibitem{leeZhang2011}
Y.-H. Lee, J.~Links and Y.-Z. Zhang,
\newblock \emph{Exact solutions for a family of spin-boson systems},
\newblock Nonlinearity \textbf{24}(7), 1975 (2011),
\newblock \doi{10.1088/0951-7715/24/7/004}.

\bibitem{chinPlenio2011}
A.~W. Chin, J.~Prior, S.~F. Huelga and M.~B. Plenio,
\newblock \emph{Generalized {{Polaron Ansatz}} for the {{Ground State}} of the
  {{Sub}}-{{Ohmic Spin}}-{{Boson Model}}: {{An Analytic Theory}} of the
  {{Localization Transition}}},
\newblock Phys. Rev. Lett. \textbf{107}(16), 160601 (2011),
\newblock \doi{10.1103/PhysRevLett.107.160601}.

\bibitem{beraFlorens2014}
S.~Bera, A.~Nazir, A.~W. Chin, H.~U. Baranger and S.~Florens,
\newblock \emph{Generalized multipolaron expansion for the spin-boson model:
  {{Environmental}} entanglement and the biased two-state system},
\newblock Phys. Rev. B \textbf{90}(7), 075110 (2014),
\newblock \doi{10.1103/PhysRevB.90.075110}.

\bibitem{divincenzoLoss2005}
D.~P. DiVincenzo and D.~Loss,
\newblock \emph{Rigorous {{Born}} approximation and beyond for the spin-boson
  model},
\newblock Phys. Rev. B \textbf{71}(3), 035318 (2005),
\newblock \doi{10.1103/PhysRevB.71.035318}.

\bibitem{luZheng2007}
Z.~L{\"u} and H.~Zheng,
\newblock \emph{Quantum dynamics of the dissipative two-state system coupled
  with a sub-{{Ohmic}} bath},
\newblock Phys. Rev. B \textbf{75}(5), 054302 (2007),
\newblock \doi{10.1103/PhysRevB.75.054302}.

\bibitem{florensNarayanan2011}
S.~Florens, A.~Freyn, D.~Venturelli and R.~Narayanan,
\newblock \emph{Dissipative spin dynamics near a quantum critical point:
  {{Numerical}} renormalization group and {{Majorana}} diagrammatics},
\newblock Phys. Rev. B \textbf{84}(15), 155110 (2011),
\newblock \doi{10.1103/PhysRevB.84.155110}.

\bibitem{pineiroorioliRey2017}
A.~Pi{\~n}eiro~Orioli, A.~{Safavi-Naini}, M.~L. Wall and A.~M. Rey,
\newblock \emph{Nonequilibrium dynamics of spin-boson models from phase-space
  methods},
\newblock Phys. Rev. A \textbf{96}(3), 033607 (2017),
\newblock \doi{10.1103/PhysRevA.96.033607}.

\bibitem{yangTong2021}
K.~Yang and N.-H. Tong,
\newblock \emph{Equilibrium dynamics of the sub-ohmic spin-boson model at
  finite temperature*},
\newblock Chinese Phys. B \textbf{30}(4), 040501 (2021),
\newblock \doi{10.1088/1674-1056/abd393}.

\bibitem{magazzuGrifoni2018}
L.~Magazz{\`u}, P.~{Forn-D{\'i}az}, R.~Belyansky, J.-L. Orgiazzi, M.~A.
  Yurtalan, M.~R. Otto, A.~Lupascu, C.~M. Wilson and M.~Grifoni,
\newblock \emph{Probing the strongly driven spin-boson model in a
  superconducting quantum circuit},
\newblock Nat Commun \textbf{9}(1), 1403 (2018),
\newblock \doi{10.1038/s41467-018-03626-w}.

\bibitem{flemingCummings2011}
C.~H. Fleming and N.~I. Cummings,
\newblock \emph{Accuracy of perturbative master equations},
\newblock Physical Review E \textbf{83}(3), 031117 (2011),
\newblock \doi{10.1103/PhysRevE.83.031117}.

\bibitem{tupkaryPurkayastha2022}
D.~Tupkary, A.~Dhar, M.~Kulkarni and A.~Purkayastha,
\newblock \emph{Fundamental limitations in {{Lindblad}} descriptions of systems
  weakly coupled to baths},
\newblock Physical Review A \textbf{105}(3), 032208 (2022),
\newblock \doi{10.1103/PhysRevA.105.032208}.

\bibitem{gilchristNielsen2005}
A.~Gilchrist, N.~K. Langford and M.~A. Nielsen,
\newblock \emph{Distance measures to compare real and ideal quantum processes},
\newblock Physical Review A \textbf{71}(6), 062310 (2005),
\newblock \doi{10.1103/PhysRevA.71.062310}.

\bibitem{yamaguchiOgawa2017}
M.~Yamaguchi, T.~Yuge and T.~Ogawa,
\newblock \emph{Markovian quantum master equation beyond adiabatic regime},
\newblock Physical Review E \textbf{95}(1), 012136 (2017),
\newblock \doi{10.1103/PhysRevE.95.012136}.

\bibitem{dimeglioHuelga2023}
G.~Di~Meglio, M.~B. Plenio and S.~F. Huelga,
\newblock \emph{Time dependent {{Markovian}} master equation beyond the
  adiabatic limit} (2023).

\bibitem{aharonovichToth2016}
I.~Aharonovich, D.~Englund and M.~Toth,
\newblock \emph{Solid-state single-photon emitters},
\newblock Nature Photonics \textbf{10}, 631 (2016),
\newblock \doi{10.1038/nphoton.2016.186}.

\bibitem{gonzalez-tudelaCirac2017}
A.~{Gonz{\'a}lez-Tudela} and J.~I. Cirac,
\newblock \emph{Quantum {{Emitters}} in {{Two}}-{{Dimensional Structured
  Reservoirs}} in the {{Nonperturbative Regime}}},
\newblock Phys. Rev. Lett. \textbf{119}(14), 143602 (2017),
\newblock \doi{10.1103/PhysRevLett.119.143602}.

\bibitem{gonzalez-tudelaCirac2017a}
A.~{Gonz{\'a}lez-Tudela} and J.~I. Cirac,
\newblock \emph{Markovian and non-{{Markovian}} dynamics of quantum emitters
  coupled to two-dimensional structured reservoirs},
\newblock Phys. Rev. A \textbf{96}(4), 043811 (2017),
\newblock \doi{10.1103/PhysRevA.96.043811}.

\bibitem{gonzalez-tudelaCirac2018}
A.~{Gonz{\'a}lez-Tudela} and J.~I. Cirac,
\newblock \emph{Non-{{Markovian Quantum Optics}} with {{Three}}-{{Dimensional
  State}}-{{Dependent Optical Lattices}}},
\newblock Quantum \textbf{2}, 97 (2018),
\newblock \doi{10.22331/q-2018-10-01-97}.

\bibitem{sheremetPoddubny2023}
A.~S. Sheremet, M.~I. Petrov, I.~V. Iorsh, A.~V. Poshakinskiy and A.~N.
  Poddubny,
\newblock \emph{Waveguide quantum electrodynamics: {{Collective}} radiance and
  photon-photon correlations},
\newblock Reviews of Modern Physics \textbf{95}(1), 015002 (2023),
\newblock \doi{10.1103/RevModPhys.95.015002}.

\bibitem{kehreinNeu1995}
S.~K. Kehrein, A.~Mielke and P.~Neu,
\newblock \emph{Flow equations for the spin-boson problem},
\newblock Zeitschrift f\"ur Physik B Condensed Matter \textbf{99}(2), 269
  (1995),
\newblock \doi{10.1007/s002570050037}.

\end{thebibliography}

\nolinenumbers

\end{document}